\documentclass[preprint,12pt,nopreprintline]{elsarticle}

\usepackage{amsmath,amssymb}
\usepackage{bm}
\usepackage{graphicx}
\usepackage{booktabs,longtable,pdflscape}
\usepackage[pass]{geometry}
\usepackage{placeins}
\usepackage{array}
\newcolumntype{L}[1]{>{\raggedright\arraybackslash}p{#1}}
\usepackage{xcolor}
\usepackage[colorlinks=true,allcolors=blue]{hyperref}

\newcommand{\tac}{t_\mathrm{ac}}
\newcommand{\pk}{p_\mathrm{k}/p_\mathrm{a}}
\begin{document}

\begin{frontmatter}

\title{Gas expansion in global models of nanosecond pulsed discharges: a reduced gas-dynamic model coupled to plasma kinetics}

\author[uowm,certh]{Konstantinos Kourtzanidis\corref{cor}}
\ead{kkourtzanidis@uowm.gr}
\author[certh]{Rention Pasolari}
\cortext[cor]{Corresponding author.}
\address[uowm]{Department of Mechanical Engineering, University of Western Macedonia,
50100 Kozani, Greece}
\address[certh]{Chemical Process \& Energy Resources Institute (CPERI), Centre for Research \&
Technology Hellas (CERTH), 6th km Charilaou-Thermi, 57001 Thessaloniki, Greece}

\begin{abstract}
Global (zero-dimensional) kinetic models of nanosecond pulsed discharges describe the expansion of the heated gas channel (kernel) with simplified gas-dynamic closures. We show that these closures can lead to errors of tens of percent in the calculated gas density, temperature and species densities, especially in the first microseconds after every pulse. Here, we propose a fast, reduced gas-dynamic model of Lagrangian shells, closed by an approximate cylindrical acoustic radiation condition, with no fitted physical coefficient. The reduced model reproduces the axis density obtained by verified one-dimensional compressible computational fluid dynamics (CFD) simulations within 2.4\% for smooth deposition profiles and axis pressures up to six times ambient, typically a thousand times faster than CFD. We couple it with our in-house global plasma-kinetics solver (SoPlasma-0D), whose energy partition after the pulse agrees well with that derived from measurements and with an independent calculation. Two microseconds after a single pulse in air, and compared with the proposed coupled model, the existing closures overestimate the axis density by up to a factor of 2.8, the atomic oxygen density by up to 2.7 and the temperature by up to 680~K. For repetitive pulses 200 or 430~ns apart, of equal energy per unit volume, no existing closure follows the coupled model at both intervals, in the gas density met by each later pulse and in the energy per molecule deposited by late pulses, which sets how strongly they dissociate and heat the gas. The proposed reduced model should therefore be used to describe the gas evolution during the expansion phase after every nanosecond pulse, and for repetitive pulses above about 100~kHz in kernels of radius up to about a millimetre.
\end{abstract}


\end{frontmatter}

\FloatBarrier
\section{Introduction}\label{sec:intro}

Nanosecond pulsed discharges have been widely studied for plasma-assisted ignition and combustion, flow control and fuel reforming \cite{PopovStarikovskaia2022}. In oxygen-containing mixtures, a large part of the deposited energy reaches the gas within tens of nanoseconds through the quenching of electronically excited states and the dissociation of molecular oxygen (fast gas heating) \cite{Popov2011,Popov2016}. The heated gas column, which we call the kernel, then expands: a pressure wave leaves the kernel and a rarefaction travels to its axis on a time of order $r/c$ (the acoustic timescale), with $r$ the kernel radius and $c$ the sound speed \cite{Xu2014}. For kernels a few hundred micrometres wide, this timescale is a few hundred nanoseconds to about a microsecond \cite{Xu2011,Xu2014,Lo2014a}. During a few acoustic times, which represent the gas expansion phase, the gas density on the axis falls, goes through a minimum and settles \cite{Xu2014}.

The expansion phase matters in two ways. First, after every pulse, whatever the repetition rate, the fast heating ends and the products of the discharge start to react within the first microseconds \cite{Popov2011,Rusterholtz2013,Lo2014a}, that is, during the expansion phase, in a gas whose density and temperature change with the expansion. Second, energy is often deposited again during the expansion phase, so that each new deposition acts on a gas whose density has been lowered by the previous ones. This happens in the secondary discharges and cable reflections that follow the main pulse in pulsed-discharge setups, about 250~ns after it on the pin-to-pin bench of Xu et al.\ \cite{Xu2011} and about 0.4~\textmu s apart in \cite{Jeong2025}, in megahertz bursts \cite{Bityurin2024}, and in double pulses with sub-microsecond delays \cite{Malla2023}. The effects of such a lowered gas density have been studied under various conditions: the gas-dynamic expansion of a streamer channel lowers the density on its axis and raises the reduced field $E/N$, which helps the transition to a spark \cite{Marode1979}, and a laser-guided discharge launched during the acoustic expansion of a filament breaks down along its low-density core \cite{Tzortzakis2001}. In microwave discharges, simulations show that gas heating lowers the gas density enough for a plasma to propagate in a field below the breakdown threshold \cite{Takahashi2017}, and that it alters the dynamics of microwave streamers \cite{Kourtzanidis2015}. In all these cases the density met by each later pulse sets its $E/N$ at a given voltage, the energy it deposits per molecule, and therefore its chemistry and heating. An accurate description of this coupling between the gas dynamics and the discharge is therefore needed to predict the effect of a discharge in practical applications.

Global (zero-dimensional) kinetic models are the usual tool for the chemistry and the energy partition of NPDs, because they carry detailed mechanisms at low cost \cite{Flitti2009,Popov2011,Lefkowitz2015,Castela2016,Pokrovskiy2022,Shao2024,Xin2024}. A global model has no spatial dimension, so it needs a closure, a prescribed law for the volume of the gas it follows. Most models keep the gas volume constant during and between pulses \cite{Flitti2009,Popov2011,Lefkowitz2015,Castela2016,Pokrovskiy2022,Xin2024}, while others integrate at constant pressure \cite{Taneja2022}, or expand the kernel isentropically to ambient pressure at the end of each pulse \cite{Shao2024}. Roger et al.\ \cite{Roger2026} derived a delayed law in which the kernel keeps its post-pulse state until the rarefaction arrives and then relaxes to the isentrope. Recently, for low-Mach computational fluid dynamics (CFD), Roger et al.\ \cite{roger2026modeling} proposed a sub-timestep model to incorporate compressibility effects, in which the authors effectively impose the state reached at the end of the expansion already during the pulse. They note that this cannot be used when the next pulse arrives before the expansion has ended.

Goutier et al.\ \cite{Goutier2026} coupled a first-order equation for the kernel volume to a global methane kinetics model. Their equation relaxes the kernel pressure on its internal acoustic time, without the inertia of the surrounding gas, so the kernel volume approaches its final value monotonically. The axis density therefore cannot fall below its final value, and a pulse that arrives near the density minimum meets a density that this model overestimates. Pai et al.\ \cite{Pai2010} used an isobaric, cycle-averaged law for the state that later pulses meet at kilohertz rates. These closures were built for, and compared with, the state between pulses or the kernel temperature and pressure. During the expansion phase the axis state goes through a minimum before it settles \cite{Xu2014}, and the error of each closure on the density met by a later pulse during this phase has not been quantified against a verified solution or a measurement.

Beyond global models, CFD simulations resolve the gas dynamics in one, two and three dimensions \cite{Xu2014,Tholin2013,Castela2016,Bityurin2024,BechaneFiorina2024}, but their computational cost grows with the dimension and with the size of the chemical mechanism, which makes them impractical for parametric studies with coupled plasma dynamics and detailed chemistry. Reduced descriptions of an expanding channel date back to Braginskii's spark-channel theory \cite{Braginskii1958}, and thin-shell models describe the transition of a cylindrical blast from the strong to the weak regime \cite{Raga2014}. These reduced descriptions give the radius of the channel or of the shock, or the state at the end of the expansion \cite{roger2026modeling}, not the density on the axis of a weakly over-pressured kernel during the expansion phase, which is the quantity a later pulse meets.

In this work we address three questions: (1) how large is the error of the existing gas-dynamic closures of global models on the gas density met by a later pulse? (2) can a reduced gas-dynamic model remove this error at a computing cost far below that of a CFD simulation? (3) what does such a model change in the predictions of a global model, after a single pulse and for a burst of pulses? To address them, we use the global (0D) solver of our in-house plasma simulation framework SoPlasma, the successor of the openly available plasma-fluid solver SoPlasmaFoam \cite{SoPlasmaFoam}, with electron-impact data from our in-house electron Boltzmann solver SoEEDF, and verified reference solutions of the compressible Euler equations computed with the CFD solvers of OpenFOAM. We then propose a reduced gas-dynamic model based on Lagrangian shells, with no physical coefficient fitted to the reference solutions, and show that it follows these solutions where the existing closures do not. The model not only needs no fitted coefficient, it also costs less than a second per case, typically a thousand times less computing time than a CFD solution. It could therefore replace the closure of a global model in two situations: in the first microseconds after every pulse, at any repetition rate, and whenever energy is deposited again during the expansion phase, by a secondary discharge, a reflection, a second pulse or a high-frequency burst.

The article is structured as follows. Section~\ref{sec:methods} defines the problem and presents the reference CFD solutions, the existing closures, the proposed reduced model, our global and electron Boltzmann solvers, the coupling of the reduced model with the kinetics, and the test cases. More details on the reduced model, including the derivation of its outer boundary condition, are given in \ref{app:boundary}. Section~\ref{sec:results} presents the results: the error of the existing closures (case~1), the verification of the reduced model (case~2), and a single pulse (case~3) and repeated pulses (case~4) computed with the coupled model. Section~\ref{sec:discussion} discusses where the model is needed, what published measurements show (they are compared with the models in \ref{app:measurements}) and the limitations of the model, and section~\ref{sec:conclusions} concludes and gives directions for future work.

\FloatBarrier
\section{Model and numerical methods}\label{sec:methods}

\FloatBarrier
\subsection{Description of the problem under study}\label{sec:problem}

We study the gas density on the axis of a kernel heated by a high-voltage nanosecond pulse, during the expansion that follows it. This density sets the rates of the chemistry that follows the pulse and, at a given applied field, the reduced field $E/N$ of any later pulse. $E/N$ in turn sets the energy that pulse deposits per molecule, its chemistry and its heating. At a fixed field $E/N$ scales as $1/N$, so a relative error $\epsilon$ in the density gives an error $1/(1+\epsilon) - 1$ in $E/N$. The quantity of interest is therefore the gas density on the axis, $\rho/\rho_0$, with $\rho_0$ the density before the first pulse, during the expansion phase and at the time a later pulse arrives.

We consider a long, straight discharge channel at mid-gap, between two electrodes. The discharge is a column with the same properties along its length, so the gas flow driven by its heating is radial: the flow is cylindrical and one-dimensional. This is the situation in the middle of pin-to-pin and pin-to-plane gaps a few millimetres long, such as those of the measurements compared in \ref{app:measurements}, and its limits are discussed in section~\ref{sec:discussion}. It is also the simplest case in which the density met by a later pulse can be computed accurately and compared between models.

The nanosecond pulse deposits its energy faster than the gas moves, so we set the time $t = 0$ at the end of the deposition, when the kernel is still at the ambient density and carries a pressure excess with the radial profile of the deposited energy:
\begin{equation}
p(r,0) = p_\mathrm{a}\left[1 + \left(\frac{p_\mathrm{k}}{p_\mathrm{a}} - 1\right) g(r)\right],
\qquad g(r) = \exp\left(-r^2/r_0^2\right),
\label{eq:initial}
\end{equation}
where $r$ is the radial coordinate, $p(r,0)$ the pressure at $t = 0$, $p_\mathrm{a}$ the ambient pressure, $p_\mathrm{k}$ the pressure on the axis at $t = 0$, $g(r)$ the radial profile of the deposited energy, equal to one on the axis, and $r_0$ its $1/e$ radius. We call $\pk$ the pressure ratio. It measures the strength of a kernel, and we use it throughout to characterise a case. For example, $\pk = 3$ is a kernel whose axis pressure is three times the ambient pressure at the end of the deposition. This instantaneous deposition is used in the comparisons with the reference CFD solutions (cases~1 and 2 of section~\ref{sec:results}). In the coupled calculations of cases~3 and 4 the pulse keeps its finite duration, 25 and 10~ns, that is 0.02 and 0.035 acoustic times (defined below). There the global solver deposits the energy over the pulse (section~\ref{sec:coupling}), and time is counted from the start of the pulse.

\begin{figure}[!htbp]
\centering
\includegraphics[width=\textwidth]{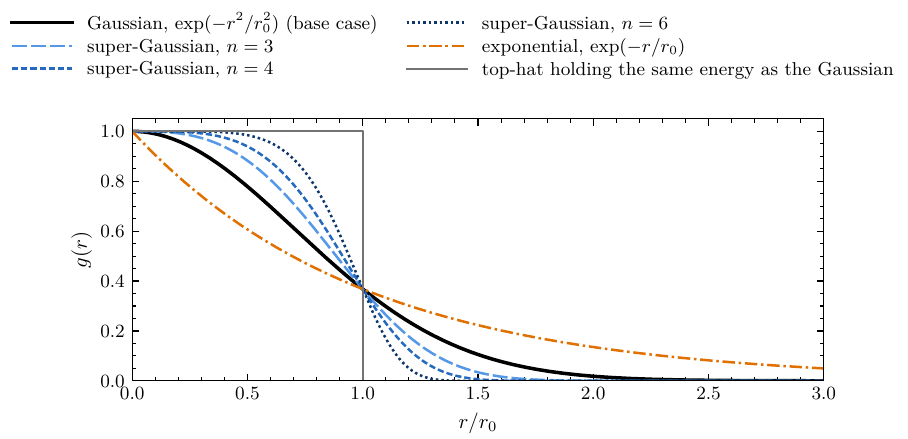}
\caption{Radial profiles $g(r)$ of the deposited energy used in this work, all with the same $r_0$: the Gaussian (base case), super-Gaussian profiles of order 3, 4 and 6, the exponential profile, and the top-hat that holds the same energy as the Gaussian.}
\label{fig:profiles}
\end{figure}

The profile of the deposited energy is an input of the problem, and it is usually assumed rather than measured. Models of discharge and laser-heated channels often take a Gaussian profile \cite{Tzortzakis2001,Xu2014}, and Ono et al.\ fitted a Gaussian profile to the measured gas density of a streamer channel \cite{Ono2010}. We use the Gaussian as the base case, and two other shapes to cover a plausible range: super-Gaussian profiles, $g(r) = \exp[-(r/r_0)^n]$ with $n > 2$, which are flatter on the axis and have a steeper edge (the profiles built on the complementary error function in \cite{Castela2016,BechaneFiorina2024} also have a steeper edge than a Gaussian), and the exponential profile $g(r) = \exp(-r/r_0)$, peaked on the axis with a long tail. Figure~\ref{fig:profiles} shows these profiles, together with the top-hat profile, uniform with a sharp edge, of the simulations behind the delayed-expansion closure (section~\ref{sec:closures}), which lies outside the scope of the reduced model (section~\ref{sec:res-verification}).

Time is scaled by the acoustic time $\tac = r_0/c_\mathrm{a}$, with $c_\mathrm{a}$ the ambient sound speed (288~ns for $r_0 = 100$~\textmu m in air at 300~K), so that a case is set by the deposition profile, the pressure ratio $\pk$ and the ratio of specific heats $\gamma$. Once the pressure on the axis has returned to the ambient pressure, the gas on the axis, heated at constant density and then expanded without exchanging heat during this time (heat conduction is slow, see section~\ref{sec:reference}), has the density
\begin{equation}
\rho_\mathrm{pl} = \rho_0 \left(\frac{p_\mathrm{a}}{p_\mathrm{k}}\right)^{1/\gamma},
\label{eq:plateau}
\end{equation}
which we call the isentropic plateau. The expansion phase is thus the temporal phase from the end of the deposition until the axis density has settled on this plateau. It lasts a few acoustic times, during which the axis density falls, goes through a minimum and recovers (as shown in section~\ref{sec:res-closures}), and ``during the expansion phase'' refers to this time window throughout. The next subsection describes the reference solution, obtained by CFD simulations, that gives the axis density of such a case.

\FloatBarrier
\subsection{Reference solution obtained by compressible CFD simulations}\label{sec:reference}

No analytical solution exists for the nonlinear expansion of a heated kernel. To measure the error of the existing closures and of the reduced model, we therefore need an accurate numerical solution of the full compressible equations, verified to a known accuracy. We compute the reference solutions with \texttt{rhoCentralFoam} of OpenFOAM v2412 \cite{Greenshields2010}, a density-based solver of the compressible Navier--Stokes equations with central-upwind fluxes, on a one-dimensional axisymmetric (wedge) mesh. We set the viscosity to zero, and with it the thermal conductivity, which the solver derives from the viscosity. The viscous stresses and the heat conduction then vanish, and the solver integrates the compressible Euler equations.
These assumptions are justified because both effects are slow during the expansion of the kernels studied here ($\pk \le 20$): heat and momentum diffuse across the kernel on times tens to hundreds of times longer than the time sound takes to cross it. The solver has been validated on flows with strong shocks \cite{Greenshields2010}. The gas is ideal with $\gamma = 1.40$ unless otherwise stated, the domain extends to 10--20 $r_0$ so that no reflection from its edge returns to the axis within the time analysed, the mesh has 100 cells per $r_0$, and the time step is adaptive with a Courant number of 0.2.

\begin{figure}[!htbp]
\centering
\includegraphics[width=\textwidth]{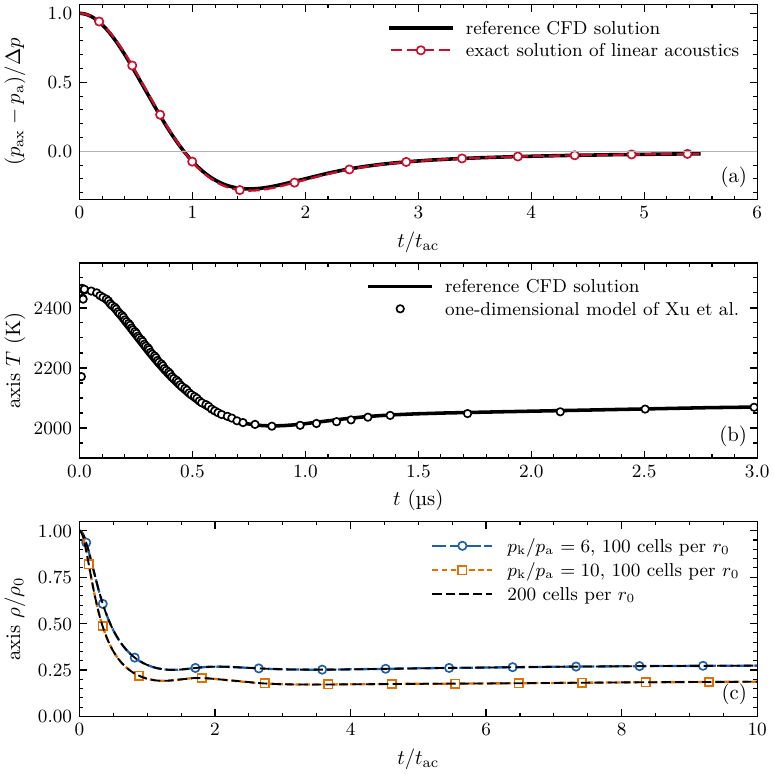}
\caption{Verification of the reference solution. (a) Axis pressure of a weak Gaussian kernel ($\pk = 1.05$, 200 cells per $r_0$) against the exact solution of linear acoustics. $\Delta p = p_\mathrm{k} - p_\mathrm{a}$. (b) Axis temperature of case~2 of Xu et al.\ \cite{Xu2014} (Gaussian kernel, $\gamma = 1.31$): one-dimensional model vs our reference solution (100 cells per $r_0$). (c) Axis density of Gaussian kernels at $\pk = 6$ and 10 with 100 and 200 cells per $r_0$.}
\label{fig:reference}
\end{figure}

We verified the CFD reference solutions in three ways and the simulation results are plotted in figure~\ref{fig:reference}. (a) For a weak Gaussian kernel ($\pk = 1.05$), whose axis pressure is known exactly from linear acoustics, the computed axis pressure departs from the exact one by at most 0.0125 of the initial pressure excess (on a mesh of 200 cells per $r_0$), in excellent agreement with linear acoustics. (b) For the one-dimensional model of Xu et al.\ \cite{Xu2014} (their case~2: a Gaussian kernel of $r_0 = 380$~\textmu m heated at constant density by 1267~K on the axis, in a gas at 1187~K, with $\gamma = 1.31$), the calculated axis temperature is in excellent agreement with the calculations of Xu et al., within 0.9\%, including the time and the value of its minimum. (c) For strong kernels no exact solution exists, so we performed a mesh convergence study: refining the mesh to 200 cells per $r_0$ changes the axis density by at most 0.1\%, also at $\pk = 6$ and 10. A later pulse, used in the repeated-pulse cases of section~\ref{sec:res-verification}, is added by raising the pressure at constant density with the profile of equation~(\ref{eq:initial}) at its current position, and the solution is restarted.

Two verification cases (section~\ref{sec:res-verification}) add a slow heat release after the deposition, to represent energy that is first stored in the vibrational reservoir and in excited states and reaches the gas at longer timescales. It is a prescribed source per unit volume, $S(r,t) = f\,E_0\,g(r)\,e^{-t/\tau}/\tau$, fixed in space, with $E_0$ the energy density deposited on the axis at $t = 0$. It adds a fraction $f$ of that energy, with the time constant $\tau$ ($f = 0.5$ with $\tau = \tac$, and $f = 1$ with $\tau = 0.3\,\tac$). We note that the developed reduced model, described in section~\ref{sec:reduced}, receives the same source, integrated over the current extent of each shell. \texttt{rhoCentralFoam} does not apply volumetric source terms, so these cases are computed with \texttt{rhoPimpleFoam}, the pressure-based compressible solver of OpenFOAM, with a semi-implicit energy source. With the source switched off, \texttt{rhoCentralFoam} and \texttt{rhoPimpleFoam} agree within 0.47\% on the axis density, and with the source on, the energy added equals the integrated source within 0.008\%.

These verified reference solutions are used to measure the error of every gas-dynamic model in section~\ref{sec:results}.

\FloatBarrier
\subsection{Existing gas dynamics closures used in global models}\label{sec:closures}

As a global model has no spatial dimension, it often replaces the gas dynamics of the kernel by a closure, a law for the volume $V(t)$ of the gas it follows, a fixed mass of gas in the kernel, from which the density follows as $\rho/\rho_0 = V_0/V$. We compare four closures, which we name as follows throughout.
\begin{enumerate}
\item The constant-volume closure: it keeps the volume, and so the density, at their values before the pulse. Most global models use it (see references in section~\ref{sec:intro}).
\item The instantaneous-expansion closure \cite{Shao2024}: it brings the kernel to the ambient pressure along the isentrope at the end of the pulse, at once, so its density jumps to the isentropic plateau $\rho_\mathrm{pl}$ of equation~(\ref{eq:plateau}) and stays there.
\item The delayed-expansion closure of Roger et al.\ \cite{Roger2026}: it holds the post-pulse state for the time $R_\mathrm{d}/c_\mathrm{k}$ that the rarefaction takes to reach the axis, with $R_\mathrm{d}$ the radius of the heated kernel and $c_\mathrm{k}$ the sound speed in the kernel. It then moves the temperature and the pressure linearly to their isentropic values, over a time that Roger et al.\ fitted to their simulations of uniformly heated (top-hat) kernels. This closure needs the kernel radius $R_\mathrm{d}$. For a Gaussian deposition we take $R_\mathrm{d} = r_0$: a top-hat of this radius with the same axis value then holds the same energy as the Gaussian, $\pi r_0^2$ per unit length in units of the axis energy density.
\item The acoustic-relaxation closure, which our global solver offers (see section~\ref{sec:solver}): it relaxes the kernel pressure towards the ambient pressure along the isentrope at every time step, with the time constant $r/c$, where $r$ is the kernel radius and $c$ the current sound speed in the kernel. Since it does not depend on when a pulse ends, it applies to any sequence of pulses. It has the same first-order form, $\mathrm{d}p/\mathrm{d}t = -(p - p_\mathrm{a})\,c/r$, as the pressure relaxation of Goutier et al.\ \cite{Goutier2026}.
\end{enumerate}
Case~1 compares all four closures with the reference CFD solution. Case~3 compares the coupled model with the first three, which describe the state after one pulse. The delayed-expansion closure is not defined when a pulse arrives during the delay or the relaxation of the previous one, so for the repeated pulses of case~4 we use the acoustic-relaxation closure in its place.

\FloatBarrier
\subsection{The proposed reduced gas-dynamic model}\label{sec:reduced}

\begin{figure}[!htbp]
\centering
\includegraphics{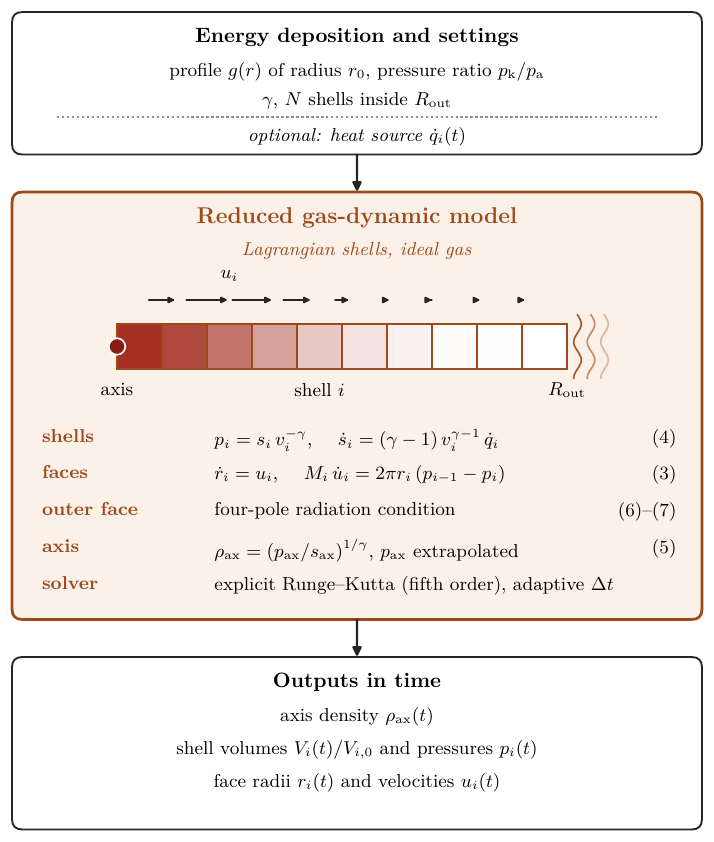}
\caption{Flow chart of the reduced gas-dynamic model: inputs, equations and outputs. The radial cut is drawn with $N = 10$ shells inside $R_\mathrm{out} = 3 r_0$, each shell coloured by the mean $g_i$ of the deposition profile over it; the arrows show the radial velocity $u_i$ of the faces during the expansion, and the outer face lets the outgoing wave leave the domain.}
\label{fig:reduced}
\end{figure}

A global model, being zero-dimensional, follows one fixed mass of gas, and the most natural choice is the gas on the discharge axis: it receives the largest energy per molecule, its density falls furthest, and it is the low-density core along which the discharge develops further \cite{Marode1979} or a later pulse meets and causes breakdown \cite{Tzortzakis2001}. The reduced model must therefore give the correct density history of the axis, at a cost close to that of a closure. It computes the whole radial structure of the expansion, but the axis is where we compare it and where it matters most for the next pulse.
To build the reduced model, we divide the gas inside a radius $R_\mathrm{out}$ into $N$ Lagrangian shells of equal initial thickness. Figure~\ref{fig:reduced} summarises the model, with its inputs, its shells and equations, and its outputs.
The shells are numbered $i = 0, \dots, N-1$ and their faces $i = 0, \dots, N$ from the axis outwards, so that shell $i$ lies between faces $i$ and $i+1$. Shell $i$ has a fixed mass per unit length $m_i$, a specific volume $v_i = \pi (r_{i+1}^2 - r_i^2)/m_i$ and an entropy variable $s_i = p_i v_i^\gamma$, so that its pressure is $p_i = s_i v_i^{-\gamma}$. At $t = 0$ the shells are at rest at the ambient density, and shell $i$ starts at the pressure $p_\mathrm{a}[1 + (\pk - 1)\,g_i]$, where $g_i$ is the mean of the deposition profile $g(r)$ of equation~(\ref{eq:initial}) over the initial cross-section of the shell. The faces obey
\begin{equation}
\frac{\mathrm{d}r_i}{\mathrm{d}t} = u_i,\qquad
M_i\,\frac{\mathrm{d}u_i}{\mathrm{d}t} = 2\pi r_i\left(p_{i-1} - p_i\right),
\label{eq:faces}
\end{equation}
with $M_i$ the mass attached to face $i$ (half of each neighbouring shell), $r_i$ and $u_i$ the radius and the radial velocity of face $i$, $t$ the time, $p_i$ the pressure of shell $i$ and $\gamma$ the ratio of specific heats of the gas. A heat source $\dot q_i$ per unit mass and per unit time enters through
\begin{equation}
\frac{\mathrm{d}s_i}{\mathrm{d}t} = (\gamma - 1)\,v_i^{\gamma - 1}\,\dot q_i,
\label{eq:entropy}
\end{equation}
and without it each shell is isentropic. No artificial viscosity is used, so a shock is carried without its entropy, which limits the model to kernels whose shocks stay weak inside $R_\mathrm{out}$ (section~\ref{sec:res-verification}). Face 0 is fixed on the axis ($r = 0$) and face $N$ is at the outer boundary. The gas on the axis is described separately, with its own entropy variable $s_\mathrm{ax}$, defined as for the shells. It starts from the state on the axis, $s_\mathrm{ax}(0) = p_\mathrm{k}/\rho_0^{\gamma}$, and changes only through a heat source, by equation~(\ref{eq:entropy}) with the heat release on the axis. Its pressure is extrapolated to $r = 0$ from the two innermost shells, quadratically in $r$, and its density follows from that pressure and its entropy:
\begin{equation}
p_\mathrm{ax} = p_0 - (p_1 - p_0)\,\frac{\bar r_0^{\,2}}{\bar r_1^{\,2} - \bar r_0^{\,2}},\qquad
\rho_\mathrm{ax} = \left(\frac{p_\mathrm{ax}}{s_\mathrm{ax}}\right)^{1/\gamma},
\label{eq:axis}
\end{equation}
with $\bar r_i^{\,2} = (r_i^2 + r_{i+1}^2)/2$, so that $\bar r_i$ divides shell $i$ into two parts of equal area. In this equation $p_0$, $p_1$, $\bar r_0$ and $\bar r_1$ belong to shells 0 and 1; they must not be confused with the initial density $\rho_0$ and the radius $r_0$ of the deposition. The axis density $\rho_\mathrm{ax}$ is the quantity that we compare with the reference solutions and that the coupled calculations pass to the global solver.

The gas outside $R_\mathrm{out}$ is not computed. It is replaced by a boundary condition on the outer face that lets the outgoing wave leave the domain without reflection, as if the gas outside extended to infinity (a radiation condition). The exact condition for a cylindrical wave depends on the whole past motion of the boundary, and we approximate it by four auxiliary ordinary differential equations (poles),
\begin{equation}
\frac{\mathrm{d}z_k}{\mathrm{d}t} = u_N - \frac{t_k c_\mathrm{a}}{R_\mathrm{out}}\,z_k,\qquad
w = u_N - \sum_{k=1}^{4} a_k\,\frac{t_k c_\mathrm{a}}{R_\mathrm{out}}\,z_k,
\label{eq:poles}
\end{equation}
with $u_N$ the velocity of the outer face, $z_k$ auxiliary variables, $w$ a filtered velocity, and $t_k$ and $a_k$ the dimensionless positions and weights of the poles: $(t_k, a_k)$ = (0.0026, 0.0068), (0.0227, 0.0459), (0.1582, 0.3042), (0.6602, 0.6431). The filtered velocity $w$ is converted into the pressure $p_\mathrm{out}$ applied on the outer face by the simple-wave relation of a polytropic gas,
\begin{equation}
p_\mathrm{out} = p_\mathrm{a}\left(1 + \frac{\gamma - 1}{2}\,\frac{w}{c_\mathrm{a}}\right)^{2\gamma/(\gamma-1)}.
\label{eq:simplewave}
\end{equation}
\ref{app:boundary} gives the derivation of this condition, a justification of its two approximations (the four-pole fit of the exact cylindrical condition and the use of a plane simple wave for a wave of finite amplitude) and of our choice of four poles. The nonlinear form of equation~(\ref{eq:simplewave}) is what keeps a domain of only three kernel radii accurate at $\pk$ = 2--3. The model has $2N + 4$ ordinary differential equations (the face positions and velocities and the four pole variables), and a heat source adds the $N + 1$ entropy variables of the shells and of the axis. With 20 shells there are only 44 equations (65 with a heat source). They are integrated with an explicit fifth-order Runge--Kutta method with adaptive time steps (relative tolerance $10^{-7}$ to $10^{-8}$, largest step $0.02\,\tac$).

The model has two numerical settings, the number of shells $N$ and the outer radius $R_\mathrm{out}$, which play the role of the mesh of a CFD solver. No coefficient of the model is fitted to make it agree with the reference solutions, and the poles approximate a known function. Like a mesh, $N$ and $R_\mathrm{out}$ must be large enough for the case. The domain must contain the deposition and keep the steepening outgoing wave inside it, so it grows for a profile with a long tail and for a strong kernel. We use $R_\mathrm{out} = 3 r_0$ with $N = 20$ shells (40 for super-Gaussian profiles and for $\pk = 3.5$), $R_\mathrm{out} = 8 r_0$ with $N = 80$ above $\pk \approx 3.5$ and for bursts, and $R_\mathrm{out} = 12 r_0$ with $N$ = 160--320 for the exponential profile, whose tail carries 20\% of the energy beyond $3 r_0$. We chose $N$ and $R_\mathrm{out}$ by refinement while comparing with the reference solutions of section~\ref{sec:reference}, taking the smallest setting that reproduced them and whose next refinement changed the axis density by less than 1\%. For a new case without a reference solution, the settings are chosen as a mesh is, by enlarging the domain and adding shells until the axis density stops changing (in \ref{app:boundary} we check the validity of this rule on our cases).

The reduced model thus gives the gas dynamics of an ideal gas for a given energy deposition and an optional heat source $\dot q_i$. In a discharge, only part of the deposited energy heats the gas during the pulse, and only part of the rest is released as heat later, at rates set by the kinetics. Sections~\ref{sec:solver} and \ref{sec:soeedf} describe the solvers that compute this kinetics, and section~\ref{sec:coupling} how their heat release enters the reduced model as an entropy history.

\FloatBarrier
\subsection{SoPlasma-0D: the global (0D) solver of SoPlasma}\label{sec:solver}

The global (0D) solver, SoPlasma-0D, was developed in-house as an integral part of SoPlasma (short for ``Simulations of Plasmas''), our in-house plasma simulation framework and the successor of the openly available plasma-fluid solver SoPlasmaFoam \cite{SoPlasmaFoam}, and shares its chemistry, thermodynamics and electron-kinetics libraries. It integrates the balance equations of all species of a mechanism,
\begin{equation}
\frac{\mathrm{d}n_i}{\mathrm{d}t} = \sum_r \left(\nu''_{ir} - \nu'_{ir}\right) k_r \prod_j n_j^{\nu'_{jr}},
\label{eq:species}
\end{equation}
together with an energy equation for the gas, an equation for the vibrational energy and, when required, an equation for the electron energy. Here $n_i$ is the density of species $i$, $\nu'_{ir}$ and $\nu''_{ir}$ are its stoichiometric coefficients as reactant and product of reaction $r$, and $k_r$ is the rate coefficient: a function of $E/N$ or of the electron mean energy for electron-impact reactions, from the electron Boltzmann solver of section~\ref{sec:soeedf}, and of the gas or electron temperature for heavy-particle reactions. The chemistry is integrated with a stiff Rosenbrock method.

The discharge can be specified by a measured profile of $E/N(t)$ with or without the electron density, by a user-defined waveform of $E/N$, or by the deposited power density. With a power-density input, $E/N$ is solved for at each step so that the requested power $P = e\,\mu_\mathrm{e} n_\mathrm{e} N (E/N)^2$ per unit volume is deposited, with $e$ the elementary charge, $\mu_\mathrm{e}$ the electron mobility, $n_\mathrm{e}$ the electron density and $N$ the gas number density. The energy requested and the energy delivered are both integrated and compared at the end of the run. The electron energy is described either in the local-field approximation, with coefficients that are functions of $E/N$, or in the local mean-energy approximation (LMEA), in which the electron energy density $n_\varepsilon = n_\mathrm{e}\bar\varepsilon$, with $\bar\varepsilon$ the mean electron energy, is integrated,
\begin{equation}
\frac{\mathrm{d}n_\varepsilon}{\mathrm{d}t} = (\mu N)\,n_\mathrm{e} N\left(\frac{E}{N}\right)^2
- n_\mathrm{e} N\,P_\mathrm{loss}(\bar\varepsilon),
\label{eq:lmea}
\end{equation}
with $\mu N$ the reduced mobility and $P_\mathrm{loss}$ the elastic and inelastic power loss per electron and per unit gas density, and the coefficients are functions of the mean energy $\bar\varepsilon$. With a power-density input the LMEA makes the inversion for $E/N$ closed-form, and it is used for all coupled calculations of this work.

The gas temperature follows
\begin{equation}
\rho c_v^\mathrm{tr}\frac{\mathrm{d}T}{\mathrm{d}t} = Q_\mathrm{el} + Q_\mathrm{def} + Q_\mathrm{heavy}
+ Q_\mathrm{VT} - Q_\mathrm{loss} - p\,\frac{1}{V}\frac{\mathrm{d}V}{\mathrm{d}t},
\label{eq:energy}
\end{equation}
where $T$ is the gas temperature, $\rho$ the mass density, $c_v^\mathrm{tr}$ the translational-rotational heat capacity at constant volume, $Q_\mathrm{el}$ the elastic heating by electrons, $Q_\mathrm{def}$ the part of the energy defect of each inelastic electron process released directly as heat, $Q_\mathrm{heavy} = \sum_r \Delta H_r R_r$ the heat of the heavy-particle reactions, $Q_\mathrm{VT}$ the vibrational-translational relaxation, $Q_\mathrm{loss}$ an optional loss, $p$ the pressure and $V$ the volume of the gas followed. In the two-temperature model the vibrational energy $e_\mathrm{v}$ of the diatomic molecules obeys
\begin{equation}
\frac{\mathrm{d}e_\mathrm{v}}{\mathrm{d}t} = Q_\mathrm{v} - Q_\mathrm{VT},\qquad
Q_\mathrm{VT} = \frac{e_\mathrm{v} - e_\mathrm{v}^\mathrm{eq}(T)}{\tau_\mathrm{VT}},
\label{eq:vib}
\end{equation}
with $Q_\mathrm{v}$ the vibrational excitation by electrons and heavy-particle reactions, $e_\mathrm{v}^\mathrm{eq}(T)$ the vibrational energy in equilibrium at the gas temperature and $\tau_\mathrm{VT}$ a Landau--Teller time with N$_2$, O$_2$ and O as collision partners \cite{Popov2011,Shao2024}. Each energy channel of equation~(\ref{eq:energy}) is integrated per unit initial volume and written out, and the solver checks at the end of every run that the channels add up to the energy delivered to the gas.

The last term of equation~(\ref{eq:energy}) is the work of expansion. The volume of the gas can be held constant (the constant-volume closure), held at constant pressure, expanded at the end of each pulse to the ambient pressure (the instantaneous-expansion closure), relaxed towards the isentrope on the time $r/c$ (the acoustic-relaxation closure), or prescribed as a history $V(t)/V_0$. In the last three, every species density and the vibrational and electron energy densities are diluted by the volume ratio, and the temperature follows the isentrope with the frozen (translational-rotational) $\gamma$ of the mixture. The prescribed history is the interface used for the coupling of section~\ref{sec:coupling}. A run can be restarted from the end state of a previous one, and the time step adapts to the imposed discharge: it is small inside each pulse, grows between pulses and lands on the start of the next one.

\begin{figure}[!htbp]
\centering
\includegraphics{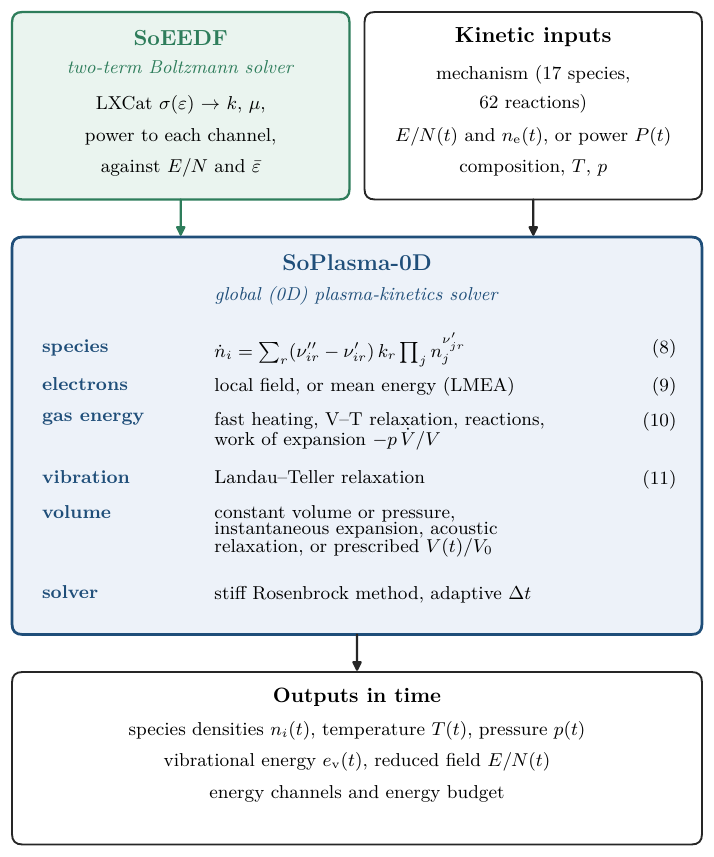}
\caption{Flow chart of our in-house global (0D) solver SoPlasma-0D and of our in-house electron Boltzmann solver SoEEDF that supplies its electron-impact data: inputs, equations and outputs. The prescribed volume history $V(t)/V_0$ is the interface with the reduced gas-dynamic model (section~\ref{sec:coupling}).}
\label{fig:solvers}
\end{figure}

SoPlasma-0D thus computes the energy partition and the heat release of a discharge, and its prescribed volume history is the interface through which the reduced model enters (section~\ref{sec:coupling}). Its electron-impact data come from the electron Boltzmann solver described next. Figure~\ref{fig:solvers} summarises the inputs, equations and outputs of the global solver and its link with the electron Boltzmann solver.

\FloatBarrier
\subsection{SoEEDF: a two-term electron Boltzmann solver}\label{sec:soeedf}

Electron-impact rate coefficients, transport coefficients and the partition of the electron power among elastic, vibrational, electronic and ionisation channels are computed by SoEEDF, a two-term electron Boltzmann solver developed in-house. It follows the formulation of BOLSIG+ \cite{HagelaarPitchford2005}, is written as a standalone C++ library with no dependency on OpenFOAM, and is embedded in our SoPlasma software. Rate coefficients and power channels are moments of the same electron energy distribution, so the rates of equation~(\ref{eq:species}) and the heat terms of equation~(\ref{eq:energy}) are consistent. The energy-space equation is discretised with an exponentially fitted scheme, the inelastic source terms are integrated exactly on piecewise-linear cross sections, superelastic collisions follow from detailed balance and the energy grid is sized automatically. The coefficients are tabulated against $E/N$ and against the mean energy, the latter for the LMEA (figure~\ref{fig:solvers}). SoEEDF reproduces BOLSIG+, the established solver of the same two-term equation: for dry air (Phelps cross sections, 79\% N$_2$ and 21\% O$_2$, $T_\mathrm{g} = 300$~K, no electron--electron or superelastic collisions, temporal growth) the mean energy, reduced mobility and reduced diffusion coefficient are in excellent agreement with BOLSIG+, within 0.03\% at all 400 reference points from 1 to 1000~Td, when both solvers use converged energy grids (figure~S1 of the supplementary material).

The air mechanism used here has 17 species and 62 reactions (24 electron-impact and 38 heavy-particle), with cross sections from the LXCat platform \cite{Pitchford2017}: the Phelps database \cite{LXCatPhelps} for N$_2$ \cite{Phelps1985} and O$_2$ \cite{Lawton1978}, and the IST-Lisbon database \cite{LXCatISTLisbon} for ground-state N \cite{Wang2014} and O (excitation and ionisation \cite{Laher1990}), and heavy-particle rates mainly from Kossyi et al.\ \cite{Kossyi1992} and Popov \cite{Popov2011,Popov2016}. The full mechanism, with the source of every rate and energy partition, is given in the supplementary material. The tables are computed for dry air at 300~K (N$_2$ 0.7885, O$_2$ 0.2115) over 0.1--2000~Td. They supply the electron-impact data of the coupled calculations, whose coupling with the reduced model is described next.

\FloatBarrier
\subsection{Coupling the reduced gas-dynamic model to the global kinetic model}\label{sec:coupling}

\begin{figure}[!htbp]
\centering
\includegraphics[width=\textwidth]{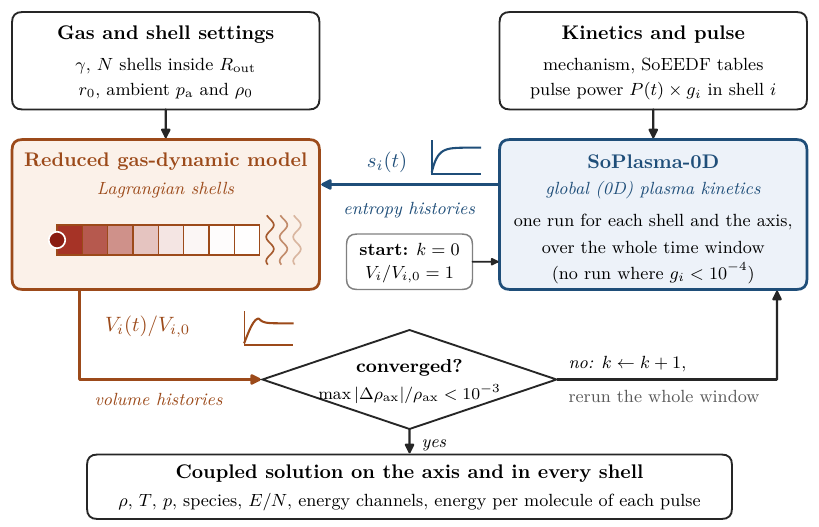}
\caption{Flow chart of the coupled model: the fixed-point iteration between the reduced gas-dynamic model and the global solver SoPlasma-0D.}
\label{fig:coupling}
\end{figure}

Here and below, the kinetics means everything that the global solver computes along a volume history: the plasma chemistry (electron-impact and heavy-particle reactions), the vibrational relaxation and the gas energy. We represent each shell of the reduced model, and the gas on the axis, by a run of the global solver with the prescribed volume history $V_i(t)/V_{i,0}$ of that shell (for the axis, $V/V_0 = \rho_0/\rho_\mathrm{ax}$; figure~\ref{fig:coupling}). The two models are solved in turn, each with the latest output of the other, until neither changes: a fixed-point iteration. We call the result of this coupling between the proposed reduced gas-dynamic model and the global kinetic model of SoPlasma the ``coupled model''.

At each iteration, the pressure $p_i(t)$ that the kinetics produces along that history is returned, and the entropy history $s_i(t) = p_i(t)\,v_i(t)^\gamma$ is passed back to the reduced gas-dynamic model, which then integrates $p_i = s_i(t)\,v_i^{-\gamma}$ with equations~(\ref{eq:faces}) and (\ref{eq:poles})--(\ref{eq:simplewave}) and returns new volume histories. The iteration starts from constant volume and stops when the largest relative change of the axis density between two iterations is below $10^{-3}$. When the fixed-point iteration has converged, the pressure of each shell equals the pressure that the kinetics produces along the shell's own path. The heat release, the change of the number of moles by dissociation and the real thermodynamics therefore enter through the entropy history, and the value of $\gamma$ in $s = p v^\gamma$ mainly affects the convergence: changing it from 1.40 to 1.45 changed the converged axis density by 0.018\%, within the iteration tolerance. In the inert outer shells, at the outer boundary and outside $R_\mathrm{out}$, $\gamma$ is that of the ambient gas.

Shell $i$ receives the power density of the pulse times $g_i$, the mean of the deposition profile over the extent of the shell, and the axis receives the full power density. For a burst, $g_i$ of each pulse is evaluated at the position of the shell at the start of that pulse, from the volume histories of the previous iteration, so that every pulse deposits the same energy density at the same place. Shells where $g_i < 10^{-4}$ would receive less than $10^{-4}$ of the energy density on the axis, too little for their chemistry to matter. They are treated as ideal gas without kinetics: they receive no energy, keep their initial entropy, and their runs of the global solver are skipped to save computing time. Each run is checked to deliver the requested energy within 0.5\%, and the time step of the global solver is fixed at 10~ps.

\begin{figure}[!htbp]
\centering
\includegraphics[width=\textwidth]{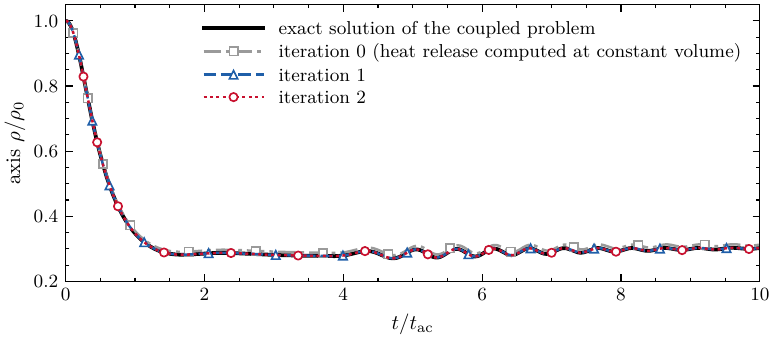}
\caption{Verification of the coupling iteration: axis density of the exact solution of the coupled problem and of the first three iterations, for a Gaussian kernel at $\pk = 3$ with all the deposited energy released after the pulse with a time constant of $0.3\,\tac$.}
\label{fig:couplingcheck}
\end{figure}

We verified the coupling algorithm, the fixed-point iteration between the reduced model and the global solver, against an exact answer. In place of the global solver and its plasma chemistry we used an ideal gas in which each shell, and the axis, receives after the pulse a prescribed heat per unit mass that depends only on time and on the shell, not on where the shell is (a heat release that follows the gas). For this test problem the coupled problem is a single system of ordinary differential equations, which we solve directly to obtain its exact solution. Figure~\ref{fig:couplingcheck} compares the axis density of the first three iterations with this exact solution. The iteration converged in two steps and reproduced that exact solution with a largest difference below 0.001\% over 10 acoustic times, while its first iterate (in which the gas dynamics is already that of the reduced model but the heat release is computed at constant volume) differed from it by 2.6--4.4\%. The test shows that the iteration has a converged solution, that it reaches this solution in only two steps, and that the volume and entropy histories exchanged by the two models are consistent with each other. We note that this verification case does not test the kinetics, which enters only in the cases of section~\ref{sec:results}.
With the plasma chemistry of section~\ref{sec:solver} the single pulse converged in two to four iterations and the bursts in five, each burst in about five minutes with five runs of the global solver computed at a time. The coupled model thus gives, for each pulse of a burst, the density, the reduced field and the energy partition, with the gas dynamics and the kinetics solved together. The next subsection defines the test cases on which we compare it (and the reduced model alone) with the reference solutions and the existing closures.

\FloatBarrier
\subsection{Test cases and models compared}\label{sec:cases}

We compare the models described above (including the new reduced gas-dynamic model) in four test cases, to measure their errors and the differences between them. Table~\ref{tab:cases} lists the test cases, which are each the subject of one subsection of section~\ref{sec:results}. Cases~1 and 2 compare the existing closures (case~1) and the reduced model (case~2) with the reference CFD solutions. Case~2 tests the reduced model in five directions: a) different radial profiles of the deposited energy, b) different pressure ratios (up to $\pk = 20$), c) a second discharge pulse 0.5 to 3 acoustic times after the first, d) bursts of five pulses and e) a slow heat release, in which part of the energy heats the gas after the pulse. Cases~3 and 4 apply the coupled model to a single pulse at the conditions of Lo et al.\ \cite{Lo2014a,Lo2014b} (a 20~mJ, 25~ns pulse in a 6.5~mm pin-to-plane gap in air at 300~K and 1~atm), and to five pulses in a 100~\textmu m kernel, 200 and 430~ns apart (5 and 2.3~MHz, that is 0.69 and 1.49 acoustic times). In cases~1 and 2 the gas is ideal and the energy is deposited at $t = 0$. In cases~3 and 4 the closures and the coupled model run in the global solver with the same plasma chemistry and the same imposed power, and differ only in the volume history of the gas. \ref{app:measurements} compares the models with two published experiments in air, as discussed in section~\ref{sec:disc-measurements}.

\begin{table}[!htbp]
\centering
\caption{Test cases, each the subject of one subsection of section~\ref{sec:results}. Cases~3 and 4 are in air at 1~atm and 300~K.}
\label{tab:cases}
\small
\begin{tabular}{L{0.9cm}L{3.0cm}L{5.0cm}L{3.2cm}}
\toprule
case & what is tested & configuration & compared with \\
\midrule
1 & existing closures & one pulse, Gaussian profile, $\pk$ = 1.1--3 & reference CFD solution \\
\addlinespace[5pt]
2 & reduced model (verification) & other profiles, $\pk$ up to 20, a second pulse, five pulses, a slow heat release & reference CFD solution \\
\addlinespace[5pt]
3 & coupled model, one pulse & conditions of Lo et al.\ \cite{Lo2014b}: Gaussian, $r_0 = 450$~\textmu m, 25~ns pulse & existing closures \\
\addlinespace[5pt]
4 & coupled model, repeated pulses & Gaussian, $r_0 = 100$~\textmu m, five 10~ns pulses 200 or 430~ns apart & existing closures \\
\bottomrule
\end{tabular}
\end{table}

For clarity, Table~\ref{tab:models} lists the gas-dynamic descriptions that these cases compare, with the names used throughout section~\ref{sec:results}. All comparisons are made on the gas density on the axis. We measure the error of a model by its relative difference from the reference CFD solution at the same time instant,
\begin{equation}
\epsilon(t) = \frac{\rho_\mathrm{m}(t) - \rho_\mathrm{ref}(t)}{\rho_\mathrm{ref}(t)},
\label{eq:error}
\end{equation}
with $\rho_\mathrm{m}$ the axis density of the model and $\rho_\mathrm{ref}$ that of the reference solution. ``The largest error'' of a case is the largest value of $|\epsilon|$ over the time window stated. In cases~3 and 4 no reference CFD solution with the plasma chemistry exists. There we compare the closures with the coupled model, whose gas dynamics case~2 verifies, and give the density of a closure relative to that of the coupled model.

\begin{table}[!htbp]
\centering
\caption{The gas-dynamic descriptions compared in this work.}
\label{tab:models}
\small
\begin{tabular}{L{5.5cm}L{7.3cm}}
\toprule
name & what it describes \\
\midrule
reference CFD solution & the compressible Euler equations on a fine one-dimensional mesh (OpenFOAM); the benchmark for every other description \\
\addlinespace[6pt]
constant-volume closure & the density stays at its value before the pulse \\
\addlinespace[6pt]
instantaneous-expansion closure & at the end of the pulse the density jumps to the isentropic plateau \\
\addlinespace[6pt]
delayed-expansion closure & the density stays at its initial value until the rarefaction reaches the axis, then relaxes to the plateau \\
\addlinespace[6pt]
acoustic-relaxation closure & the pressure relaxes to ambient along the isentrope with the time constant $r/c$ \\
\addlinespace[6pt]
reduced model & the model proposed here based on Lagrangian shells, for a prescribed energy deposition \\
\addlinespace[6pt]
coupled model & the reduced model coupled to the global solver, which computes the heat release with the plasma chemistry \\
\bottomrule
\end{tabular}
\end{table}

\FloatBarrier
\section{Results}\label{sec:results}

\FloatBarrier
\subsection{Case 1: Error of the existing closures during the expansion phase}\label{sec:res-closures}

\begin{figure}[!htbp]
\centering
\includegraphics[width=\textwidth]{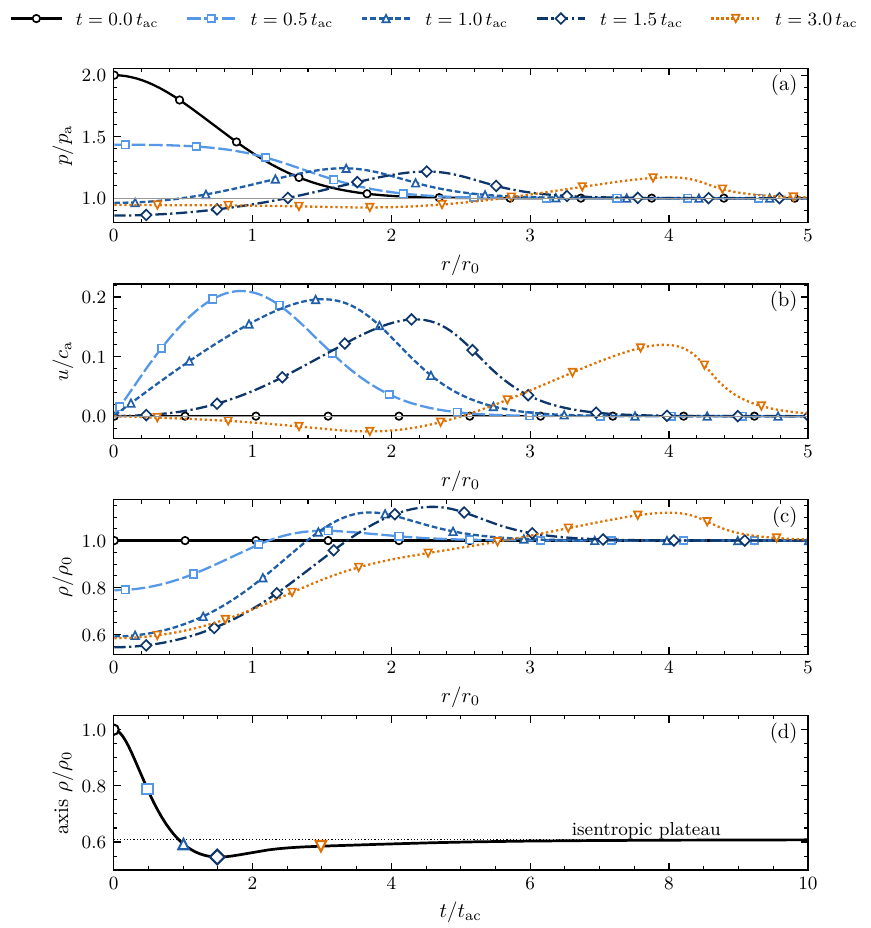}
\caption{Case 1, reference CFD solution for a Gaussian kernel at $\pk = 2$. (a) Pressure, (b) radial velocity and (c) density against radius at five times; (d) the axis density against time, with the five times marked and the isentropic plateau (dotted).}
\label{fig:anatomy}
\end{figure}

\begin{figure}[!htbp]
\centering
\includegraphics[width=\textwidth]{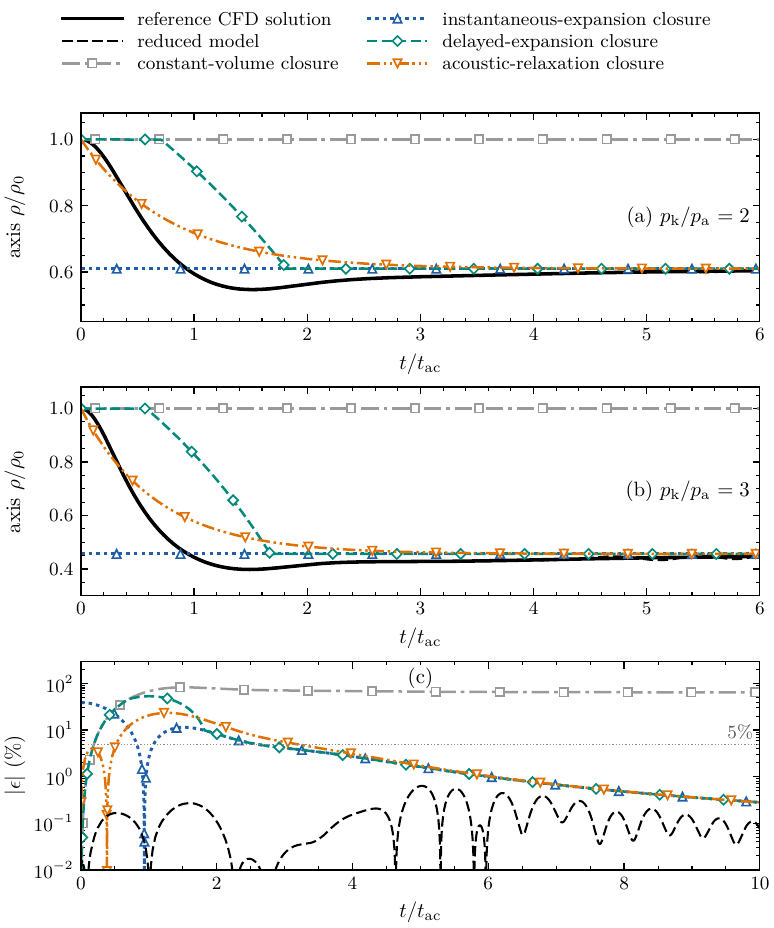}
\caption{Case 1, axis density after one pulse of a Gaussian kernel at $\pk = 2$ (a) and 3 (b): the reference CFD solution, the four existing closures, and the reduced model ($R_\mathrm{out} = 3 r_0$, $N = 20$; discussed in case 2). (c) Magnitude of the error $|\epsilon|$ (equation~(\ref{eq:error})) of each closure and of the reduced model at $\pk = 2$, on a logarithmic scale; the dotted line marks 5\%.}
\label{fig:closures}
\end{figure}

 We first compare the four closures of section~\ref{sec:closures} with the reference CFD solution for a single Gaussian kernel heated at constant density, at pressure ratios $\pk$ from 1.1 to 3, with $\gamma = 1.40$. The acoustic-relaxation closure takes $r = r_0$. The closures start from the same state as the reference, so any difference comes from their form. We start with the expansion in the reference solution (figure~\ref{fig:anatomy}) and then compare the closures with it (figure~\ref{fig:closures} and table~\ref{tab:closures}).

Figure~\ref{fig:anatomy} plots, for $\pk = 2$, the pressure (a), the radial velocity (b) and the density (c) against radius at five times after the deposition, and the axis density against time (d), with the five times marked. The pressure excess pushes the gas outwards (b): a compression wave leaves the kernel and carries energy away, and a rarefaction travels inwards and reaches the axis within about one acoustic time, so the axis density falls (c, d). When the axis pressure has returned to ambient, at about $0.9\,\tac$, the gas around the axis is still moving outwards, and its inertia carries the axis density below the isentropic plateau. The pressure near the axis, now below ambient, stops this motion at $1.5\,\tac$, when the axis density reaches its minimum, about 10\% below the plateau (d). The axis density then recovers slowly towards the plateau.

Figure~\ref{fig:closures} compares the axis density of the reference solution with that of the four closures at $\pk = 2$ (a) and 3 (b), and plots the magnitude of the error $|\epsilon|$ of each closure at $\pk = 2$ on a logarithmic scale (c). The figure also shows the reduced model, which is discussed in case~2 below. Table~\ref{tab:closures} lists the error of each closure at every pressure ratio at three times, 0.5, 1.0 and $1.5\,\tac$ (the time of the density minimum).

The reference solution has the history of figure~\ref{fig:anatomy} at every pressure ratio from 1.1 to 3. The axis density has fallen by 5\% after 0.13 to $0.7\,\tac$, the earlier the stronger the kernel. It reaches its minimum at $1.5\,\tac$ at every pressure ratio, 2 to 13\% below the plateau, and it returns to within 5\% of the plateau after $2.5\,\tac$ at $\pk = 2$ and $4\,\tac$ at $\pk = 3$. The time of the minimum is therefore set by the width of the deposition, through the acoustic time, and its depth by the deposited energy, through the pressure ratio. These few acoustic times are the expansion phase defined in section~\ref{sec:problem}.

None of the closures follows this history (figure~\ref{fig:closures}, table~\ref{tab:closures}). The constant-volume closure keeps the initial density, so its error grows as soon as the rarefaction reaches the axis, to $+151\%$ at the density minimum for $\pk = 3$. Since its density never reaches the plateau, the error remains after the expansion phase. The instantaneous-expansion closure jumps to the plateau at once, so it is too low before the minimum and too high around it ($-30\%$ at $0.5\,\tac$ and $+15\%$ at $1.5\,\tac$ for $\pk = 3$). The delayed-expansion closure holds the initial density too long and reaches the plateau too late ($+87\%$ at one acoustic time for $\pk = 3$). The acoustic-relaxation closure follows the early fall best ($+9\%$ at $0.5\,\tac$ for $\pk = 3$), but it approaches the plateau monotonically and never falls below it, so it is too dense around the minimum ($+30\%$ at one acoustic time for $\pk = 3$). From $\pk = 1.3$ to 3, no closure stays within 5\% of the reference over the whole window from 0.2 to $2.5\,\tac$ (figure~\ref{fig:closures}(c)); at $\pk = 1.1$ only the acoustic-relaxation closure does (3.6\% at most). Section~\ref{sec:discussion} gives this window in microseconds for several kernels. At a fixed applied field these errors pass into the reduced field of a later pulse, since $E/N$ scales as $1/N$: at $\pk = 3$ the reduced field of a later pulse is 60\% too low with the constant-volume closure at the density minimum, and 43\% too high with the instantaneous-expansion closure at $0.5\,\tac$.

\begin{table}[!htbp]
\centering
\caption{Case 1, error $\epsilon$ (equation~(\ref{eq:error}), in \%) of the existing closures on the axis density at 0.5 / 1.0 / $1.5\,\tac$ after the deposition ($1.5\,\tac$ is the time of the density minimum), Gaussian kernel, $\gamma = 1.40$. A positive value means that the closure is too dense. The delayed-expansion closure takes $R_\mathrm{d} = r_0$ and the acoustic-relaxation closure $r = r_0$.}
\label{tab:closures}
\small
\setlength{\tabcolsep}{4pt}
\begin{tabular}{ccccc}
\toprule
 & constant & instantaneous & delayed & acoustic \\
$\pk$ & volume & expansion & expansion & relaxation \\
\midrule
1.1 & $+3$/$+8$/$+9$ & $-4$/$+0.5$/$+2$ & $+3$/$+7$/$+5$ & $+0.3$/$+3$/$+3$ \\
1.3 & $+9$/$+22$/$+27$ & $-10$/$+1$/$+5$ & $+9$/$+20$/$+14$ & $+1$/$+8$/$+9$ \\
1.5 & $+15$/$+36$/$+44$ & $-14$/$+2$/$+7$ & $+15$/$+31$/$+22$ & $+2$/$+13$/$+14$ \\
2 & $+28$/$+68$/$+83$ & $-22$/$+3$/$+12$ & $+28$/$+53$/$+35$ & $+4$/$+21$/$+22$ \\
3 & $+53$/$+125$/$+151$ & $-30$/$+3$/$+15$ & $+53$/$+87$/$+43$ & $+9$/$+30$/$+29$ \\
\bottomrule
\end{tabular}
\end{table}

The existing closures therefore miss the axis density by tens of percent during the expansion phase, because none of them represents both the duration of the expansion and the fall below the plateau. These errors come from the form of the closures, not from their inputs, which are those of the reference.

\FloatBarrier
\subsection{Case 2: Verification of the reduced model}\label{sec:res-verification}

 In case~2 we ask whether the reduced model reproduces the axis density of the reference CFD solution, also in cases that were not used to build it. While developing the reduced model, which has no fitted coefficient, we compared it only with the five Gaussian cases of case~1, with $N = 20$ shells inside $R_\mathrm{out} = 3 r_0$. Figure~\ref{fig:closures} already shows two of them ($\pk = 2$ and 3). For the other cases we did not change the model: we only increased the number of shells $N$ and the outer radius $R_\mathrm{out}$ where needed, as one refines a mesh, by the rule of section~\ref{sec:reduced} (\ref{app:boundary} shows that this refinement test alone detects a domain that is too small). We compared the reduced model with the reference in five other kinds of cases: other deposition profiles, larger pressure ratios, a second pulse, bursts of five pulses and a slow heat release (as described in section~\ref{sec:cases}). For every case we report the largest error $|\epsilon|$ over the computed window, which extends to $10\,\tac$ after the last pulse.

\begin{figure}[!htbp]
\centering
\includegraphics[width=\textwidth]{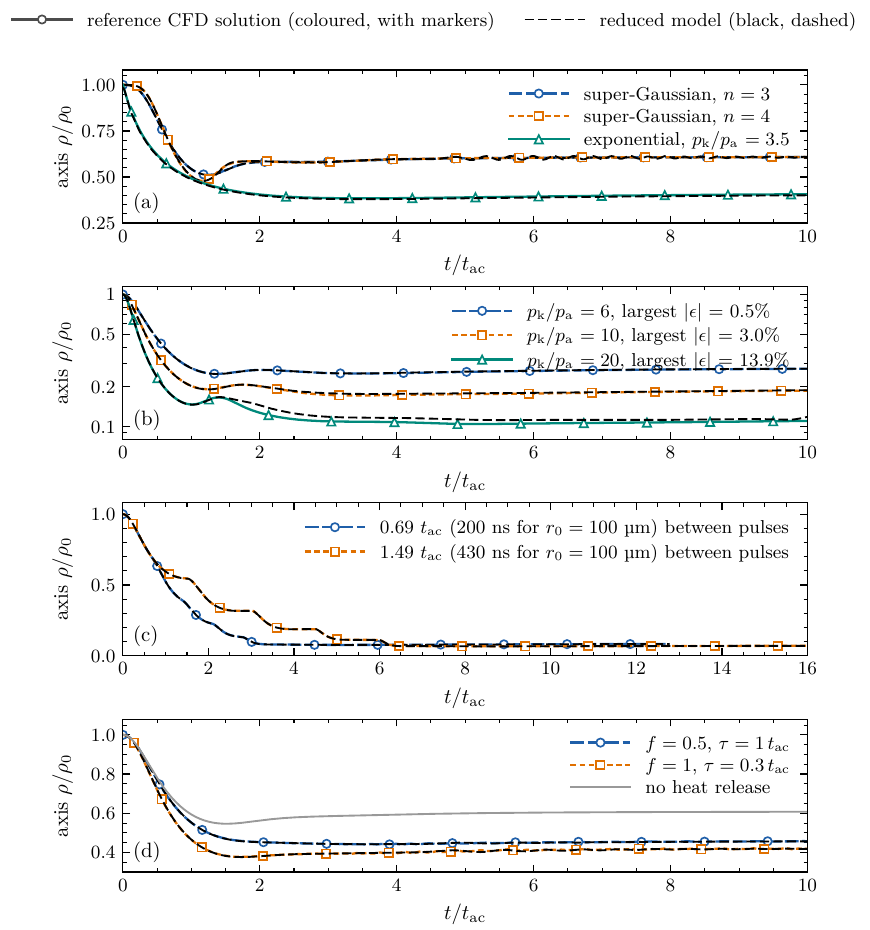}
\caption{Case 2, axis density of the reduced model (dashed black lines) and of the reference CFD solution (solid coloured lines): (a) super-Gaussian profiles of order $n = 3$ and 4 at $\pk = 2$, and an exponential profile at $\pk = 3.5$; (b) Gaussian kernels at $\pk$ = 6, 10 and 20, on a logarithmic scale, with the largest error $|\epsilon|$ of each case in the legend; (c) bursts of five pulses with 0.69 and $1.49\,\tac$ between pulses (200 and 430~ns for $r_0 = 100$~\textmu m); (d) a slow heat release of a fraction $f$ of the energy deposited at $t = 0$, with the time constant $\tau$ (section~\ref{sec:reference}), and the reference solution without heat release (grey).}
\label{fig:verification}
\end{figure}

Figure~\ref{fig:verification} plots the axis density of the reduced model (dashed black lines) over that of the reference solution (solid coloured lines) for four groups of cases: (a) super-Gaussian profiles of order 3 and 4 at $\pk = 2$ and an exponential profile at $\pk = 3.5$; (b) Gaussian kernels at $\pk = 6$, 10 and 20, on a logarithmic scale, so that the low densities reached at late times can be compared, with the largest error of each case in the legend; (c) bursts of five pulses with 0.69 and $1.49\,\tac$ between pulses; and (d) a slow heat release, with the reference solution without heat release for comparison. Where the two lines overlap, the reduced model follows the reference. Table~\ref{tab:verification} lists the largest error of every case, with the settings used.

On the five Gaussian cases of case~1 the largest error of the reduced model is 0.6\% up to $\pk = 2$ and 2.2\% at $\pk = 3$ (figure~\ref{fig:closures}(c)), against tens of percent for the closures. On the cases of the other kinds it stays within 2.4\%: for super-Gaussian profiles up to order 4 and the exponential profile, for Gaussian kernels up to $\pk = 6$, for a second pulse 0.5 to 3 acoustic times after the first, for bursts of five pulses in which the axis density falls below 0.08 of its initial value, and for a slow heat release that changes the reference axis density by up to a third. At $\pk = 10$ the largest error is 3.0\%.

\begin{table}[!htbp]
\centering
\caption{Case 2, largest error $|\epsilon|$ of the reduced model on the axis density against the reference CFD solution, over the computed window with the settings used ($N$ shells inside $R_\mathrm{out}$).}
\label{tab:verification}
\small
\begin{tabular}{L{6.45cm}cL{3.75cm}}
\toprule
case & $R_\mathrm{out}/r_0$, $N$ & largest error \\
\midrule
\multicolumn{3}{l}{\textbf{Gaussian profile, one pulse}} \\
\hspace{1em}$\pk$ = 1.1, 1.3, 1.5, 2, 3 & 3, 20 & 0.0, 0.1, 0.3, 0.6, 2.2\% \\
\hspace{1em}$\pk = 3.5$ & 3, 40 & 1.8\% \\
\hspace{1em}$\pk$ = 6, 10 & 8, 80 & 0.54, 3.0\% \\
\hspace{1em}$\gamma = 1.31$, $\pk = 2.06$ (Xu et al.\ \cite{Xu2014}) & 3, 20 & 0.7\% \\
\midrule
\multicolumn{3}{l}{\textbf{Other profiles, one pulse}} \\
\hspace{1em}super-Gaussian, $n$ = 3, 4, 6, $\pk = 2$ & 3, 40 & 0.8, 1.8, 6.3\% \\
\hspace{1em}exponential, $\pk$ = 3.5, 10 & 12, 160 / 320 & 1.7, 1.4\% \\
\midrule
\multicolumn{3}{l}{\textbf{Repeated pulses and slow heat release (Gaussian profile)}} \\
\hspace{1em}second pulse, $\pk = 2$, at 0.49, 1.01, 1.49, 2.99 $\tac$ & 3, 20 & 1.1, 0.7, 1.2, 1.4\% \\
\hspace{1em}five pulses, each raising the axis pressure by $p_\mathrm{a}$, 0.69 and 1.49 $\tac$ apart & 8, 80 & 0.92, 0.88\% \\
\hspace{1em}slow heat release, $\pk = 2$ & 3, 20 & 1.0, 2.4\% \\
\bottomrule
\end{tabular}
\end{table}

The settings behave like a mesh. In our refinement tests the error roughly halves when the number of shells doubles, until it levels off at 0.5--1.5\% (Gaussian at $\pk = 6$, exponential profile), an error that comes from the outer boundary condition and not from the number of shells. Profiles with a steeper edge need more shells (40 for a super-Gaussian of order 4, while order 6 is still 6.3\% off with 40 and would need more, which we did not test), and stronger kernels need a larger domain (at $\pk = 6$, 8.4\% with $R_\mathrm{out} = 3 r_0$ and 0.54\% with $8 r_0$).

Two kinds of conditions lie outside the scope of the model, and neither occurs in the discharges compared here. (a) With a top-hat profile, uniform with a sharp edge, the rarefaction reaches the axis from all sides at the same moment (it focuses). The axis density then changes so sharply that even the reference solution does not converge with the mesh at that moment. There the reduced model is 74\% off, and up to 33\% at the other times. We note that the deposition profiles assumed in models of nanosecond discharges, and the Gaussian fitted to a measured streamer channel, have an edge of finite width (section~\ref{sec:problem}): the top-hat is an idealisation. (b) Kernels at $\pk = 20$ or more launch (strong) shocks. A shock raises the entropy of the gas it crosses, but the shells of the reduced model stay isentropic and miss this rise. On the logarithmic scale of figure~\ref{fig:verification}(b) the two curves at $\pk = 20$ look close, but the error reaches 14\% at about two acoustic times, when the axis density of the reference falls a second time. In the experiments compared in \ref{app:measurements} the kernels lie at $\pk \approx 2$ (Rusterholtz et al.) and 3 (Lo et al.), inside the range verified here. Thermal sparks, for which Goutier et al.\ \cite{Goutier2026} compute 11--33~bar, lie beyond the validity range of our model.

For smooth profiles (Gaussian, exponential and super-Gaussian up to order 4) and $\pk$ up to 6, the reduced model is in good agreement with the reference CFD solution, within 2.4\% on the axis density (3.0\% at $\pk = 10$), in 0.05--0.3~s of computing time per case, against one to fifteen minutes on one core for the reference CFD solution: typically a thousand times less. This makes it usable inside a global kinetic model, which we do next.

\FloatBarrier
\subsection{Case 3: A single pulse with the coupled model}\label{sec:res-single}

In case~3 we ask what the description of the gas dynamics changes in the predictions of a global model after a single pulse, once the plasma chemistry is included. We couple the reduced model to the global solver (the coupled model, section~\ref{sec:coupling}) and compare it with the constant-volume, instantaneous-expansion and delayed-expansion closures. All four runs use the same plasma chemistry and the same imposed power density, so they differ only in the volume history of the gas. We take the conditions of the pin-to-plane discharge of Lo et al.\ \cite{Lo2014a,Lo2014b}: dry air at 300~K and 1~atm, a 25~ns pulse, and a Gaussian deposition with $r_0 = 450$~\textmu m ($\tac = 1.3$~\textmu s), the width that fits the radial profiles they measured at 150~ns. The energy deposited on the axis was not measured. We set it to $3.07\times10^{6}$~J\,m$^{-3}$ (0.78~eV per molecule), so that the coupled model gives the axis pressure measured at 150~ns. This fixes an unmeasured input of the case, and nothing in the model is adjusted. At each iteration of the coupling the global solver is run once for each of the 40 shells (inside $R_\mathrm{out} = 3 r_0$) and once for the axis, and the coupled solution converged in two to four iterations.

\begin{figure}[!htbp]
\centering
\includegraphics[width=\textwidth]{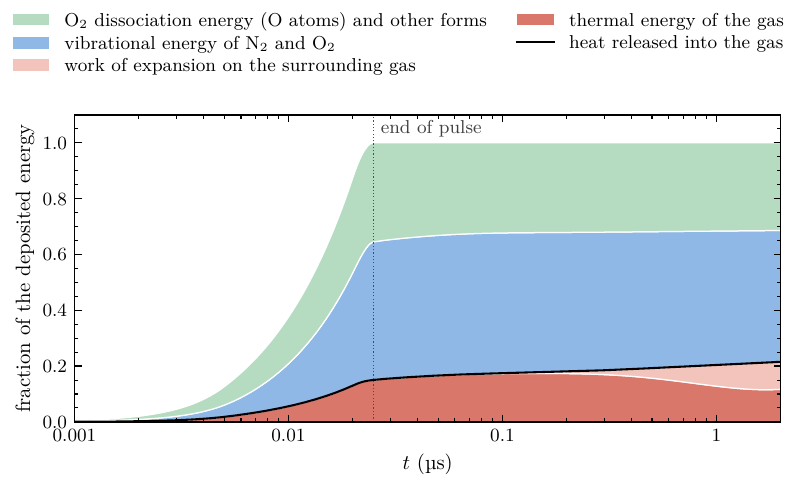}
\caption{Case 3, energy partition on the axis in the coupled model, for one pulse at the conditions of Lo et al.: thermal energy of the gas, work of expansion, vibrational energy, and O$_2$ dissociation energy and other forms, as fractions of the energy deposited by the end of the pulse (25~ns).}
\label{fig:energy}
\end{figure}

We first look at where the deposited energy goes, because only the part that heats the gas raises the pressure that drives the expansion. The electrons spend the energy of the pulse mainly on the vibrational excitation of N$_2$, the electronic excitation of N$_2$ and O$_2$, the dissociation of O$_2$ and ionisation. The part of this energy that heats the gas within tens of nanoseconds is the fast gas heating. It comes mainly from the quenching of electronically excited N$_2$ by O$_2$, which dissociates O$_2$, from the quenching of O($^1$D) by N$_2$, and from the dissociation of O$_2$ by electron impact, whose fragments carry kinetic energy \cite{PopovStarikovskaia2022}. In the global solver it is given by the terms $Q_\mathrm{def}$ and $Q_\mathrm{heavy}$ of equation~(\ref{eq:energy}). The energy stored in the vibration of N$_2$ returns as heat much later, through vibrational-translational (V-T) relaxation (the slow gas heating), and the energy stored in the O atoms is released only when they recombine or react. Figure~\ref{fig:energy} shows the energy partition on the axis in the coupled model, that is, how the deposited energy is shared at each time between four parts that add up to it: a) the thermal energy of the gas and b) the work of expansion done by the gas on the surrounding gas, which together are the heat released into the gas (black line); c) the vibrational energy of N$_2$ and O$_2$ and d) the rest, held mainly as the dissociation energy of O$_2$ (in the O atoms), with a few percent in excited states of O$_2$ and in forms that our kinetics does not resolve. The fractions shown are relative to the energy deposited by the end of the pulse.

By the end of the pulse (25~ns), processes responsible for fast gas heating have released 15\% of the deposited energy as heat, 4.6\% through the energy defects of electron-impact processes and 10.3\% through heavy-particle reactions. About half of the deposited energy (49\%) is stored as vibrational energy of N$_2$ and O$_2$ and about a third in other forms, most of it (24\% of the deposited energy) as the dissociation energy of O$_2$ held by the O atoms. Table~\ref{tab:energy} gives our split at 150~ns, the time of the first measurements of Lo et al., and compares it with the split that Lo et al.\ derived from their measured gas temperature, vibrational distribution and composition \cite{Lo2014b}, and with the split that Popov computed on the axis of the same discharge \cite{Popov2016}. Our split is in good agreement with both, since each of our fractions differs from theirs by at most 5\% of the deposited energy. In all three, about a sixth of the deposited energy has heated the gas, about half is stored as vibrational energy, and a fifth to a quarter as the dissociation energy of the O atoms. Note that Lo et al.\ give their split for the whole channel (assumed uniform along its length), and it accounts for 87.5\% of the energy they measured, while Popov and this work give it on the axis; our vibrational energy also includes that of O$_2$. The agreement therefore supports the overall energy partition of our kinetics, but it does not test each reaction separately.

\begin{table}[!htbp]
\centering
\caption{Case 3, partition of the deposited energy at 150~ns after one pulse at the conditions of Lo et al., in percent of the deposited energy (see text).}
\label{tab:energy}
\small
\begin{tabular}{L{5.2cm}ccc}
\toprule
 & heat & vibration & O$_2$ dissociation \\
\midrule
Lo et al.\ \cite{Lo2014b} (measurements) & 18.7 & 47.5 (N$_2$) & 20 \\
Popov \cite{Popov2016} (calculation) & 17 & 55 (N$_2$) & 27 \\
this work (coupled model) & 17.8 & 50.1 (N$_2$, O$_2$) & 25.0 \\
\bottomrule
\end{tabular}
\end{table}

From the end of the pulse to 2~\textmu s the heat released grows only from 15\% to 21.5\% of the deposited energy: V-T relaxation adds 3.6\% and reactions about 3\%. By 2~\textmu s almost half of this heat (9.9\% of the deposited energy) has been converted into work of expansion on the surrounding gas. The expansion is therefore driven by about a fifth of the deposited energy, released mostly by the fast gas heating during the pulse and partly during the expansion phase. Both the amount and the timing of this heat are outputs of the plasma chemistry, which is why in our coupled model the gas dynamics is solved together with the kinetics.

\begin{figure}[!htbp]
\centering
\includegraphics[width=\textwidth]{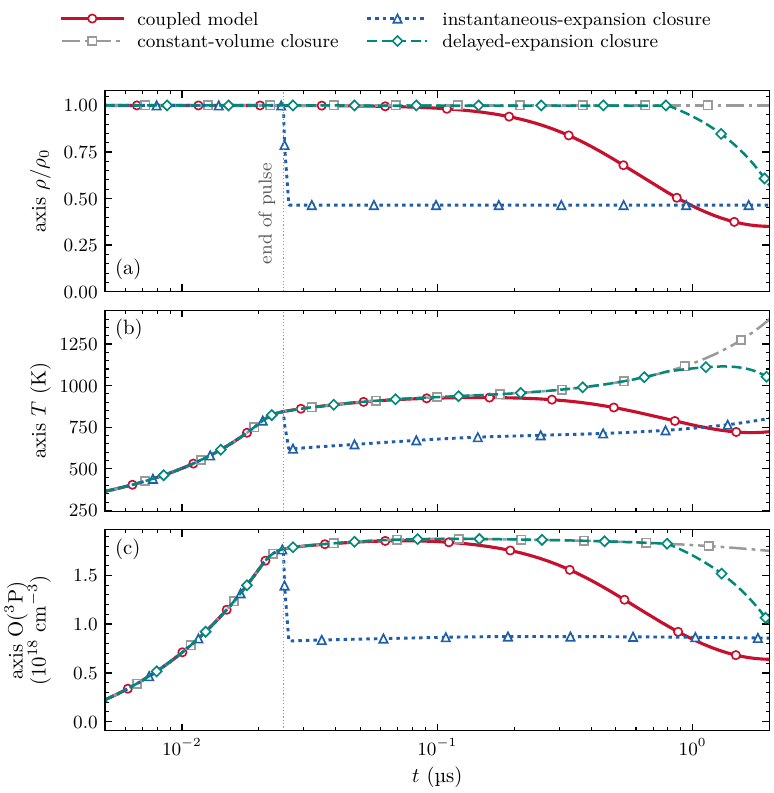}
\caption{Case 3, one pulse at the conditions of Lo et al.\ \cite{Lo2014a,Lo2014b}, on the axis: (a) gas density, (b) gas temperature and (c) O($^3$P) density of the coupled model and of the three existing closures. The dotted vertical line marks the end of the pulse.}
\label{fig:single}
\end{figure}

Figure~\ref{fig:single} compares, on the axis and over two microseconds, the gas density, the gas temperature and the O($^3$P) density of the coupled model with those of the three closures. The results are identical up to the end of the pulse, since the gas does not move during it, and then separate: the instantaneous-expansion closure expands at once, too early, the delayed-expansion closure too late (after 0.77~\textmu s), and the constant-volume closure never. At 2~\textmu s, about 1.5 acoustic times for this kernel, the density of the coupled model is still falling slowly, and the closures give 1.3 times (instantaneous expansion) to 2.8 times (constant volume) the density of the coupled model. The O($^3$P) density follows the gas density, because every species is diluted by nearly the same factor. The O($^3$P) mole fraction differs by less than 4\% between the four runs, and the constant-volume closure gives 2.7 times the O($^3$P) density of the coupled model. The closures give temperatures up to 680~K above that of the coupled model. Two effects give the constant-volume closure the highest temperature: (a) it does no work of expansion, so it keeps as heat the 9.9\% of the deposited energy that the coupled model spends on expansion (figure~\ref{fig:energy}), and (b) its vibrational relaxation, which is faster in a dense and hot gas, returns 11.5\% of the deposited energy as heat instead of 3.6\%.

It is important to note that this result concerns every pulse, whether or not other pulses follow. A global model is used to predict what species a pulse produces: atomic oxygen, electronically and vibrationally excited molecules, and the products of their reactions. These species react during the first microseconds in gas conditions that each closure describes, and the rates of the three-body reactions and of the vibrational relaxation depend on the gas density and the temperature. Errors of tens of percent in both during this period therefore carry into the gas densities and the temperature from which the afterglow starts. With the constant-volume closure the error does not even vanish after the expansion phase, since its density never reaches the plateau. Here we run simulations with the coupled model for two microseconds. We note though that over longer times, heat conduction and the induced flows also change the density on the axis (see also section~\ref{sec:discussion}). Beyond single pulses, when energy is deposited again during the expansion phase, later pulses might be strongly influenced by the gas state left from previous ones, which case~4 addresses below.

\FloatBarrier
\subsection{Case 4: Repeated pulses during the expansion phase}\label{sec:res-burst}

In case~4 we ask what happens when energy is deposited again during the expansion phase, so that each pulse meets the gas left by the previous ones. We compute five 10~ns pulses in a kernel of $r_0 = 100$~\textmu m ($\tac = 288$~ns), with a time period of 200~ns between pulses (5~MHz, $0.69\,\tac$) and also with 430~ns (2.3~MHz, $1.49\,\tac$, so that each pulse arrives near the density minimum left by the previous one). We compare our coupled model with the constant-volume, instantaneous-expansion and acoustic-relaxation closures (section~\ref{sec:closures}). All simulations use the same plasma chemistry and the same imposed power. We assume, for this demonstration, that every pulse deposits $5\times10^{5}$~J\,m$^{-3}$ on the axis (0.128~eV per molecule in the first pulse) with the same radial profile, fixed in space.%
Under these conditions, the axis pressure stays below 1.4 times the ambient pressure throughout, inside the verified range of the reduced model (case~2).

\begin{figure}[!htbp]
\centering
\includegraphics[width=\textwidth]{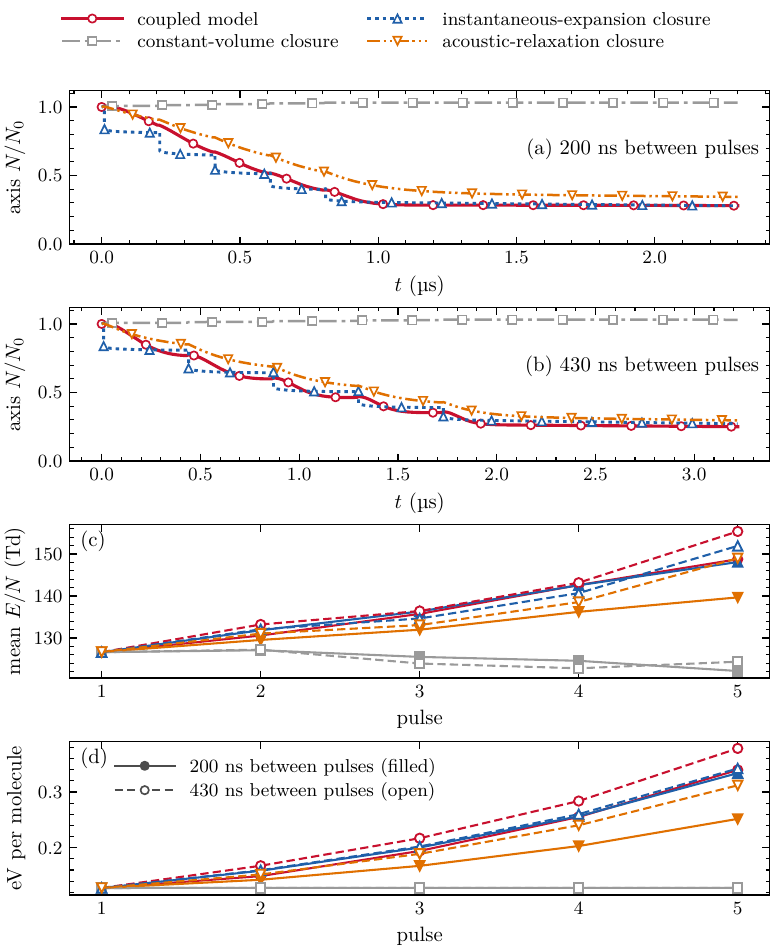}
\caption{Case 4, five 10~ns pulses in a kernel of $r_0 = 100$~\textmu m. Axis number density against time with 200~ns (a) and 430~ns (b) between pulses; (c) the mean reduced field during each pulse and (d) the energy deposited per molecule in each pulse (solid lines and filled markers: 200~ns; dashed lines and open markers: 430~ns).}
\label{fig:burst}
\end{figure}

Figure~\ref{fig:burst} plots the axis number density $N/N_0$ of the four models
against time with 200~ns and 430~ns between pulses, and, for each pulse, the mean reduced field during the pulse and the energy deposited per molecule. The energy per molecule of a pulse is the energy it deposits per unit mass of the gas on the axis, expressed in electronvolts per molecule of air (that is, multiplied by the mean mass of an air molecule). It depends only on the mass density, so the dissociation of O$_2$ does not change it. The number density $N$ of figure~\ref{fig:burst} and table~\ref{tab:burst} counts all particles, atoms included, so at constant volume it rises slightly, to $1.027\,N_0$ at the fifth pulse. The first pulse deposits 0.128~eV per molecule in every model. Since every pulse deposits the same energy per unit volume, the energy per molecule of a later pulse grows in the same ratio as the gas has expanded. Table~\ref{tab:burst} lists the density ``seen'' by each pulse and the energy per molecule deposited by the fifth pulse.

In the coupled model each pulse meets a lower density than the previous one, down to 0.39 (200~ns) and 0.35 (430~ns) of the initial density at the fifth pulse. Note that we impose the power density, not the applied voltage. The global solver then finds the $E/N$ that deposits this power, and this $E/N$, set by the plasma chemistry, rises only from 127 to about 150~Td. The lower density therefore shows mainly in the energy deposited per molecule, which grows by factors of 2.7 and 3.0 from the first to the fifth pulse (figure~\ref{fig:burst}(d)). The deposited energy per molecule sets how strongly each pulse dissociates and heats the gas.%

\begin{table}[!htbp]
\centering
\caption{Case 4, axis number density ``seen'' by pulses 2 to 5, relative to its value before the first pulse, and energy deposited per molecule by pulse 5, in eV (pulse 1 deposits 0.128~eV per molecule in every model). $N/N_0$ counts the atoms produced by dissociation, so it is up to 4\% above the ratio of the mass densities. }
\label{tab:burst}
\resizebox{\textwidth}{!}{%
\begin{tabular}{lccccc}
\toprule
 & \multicolumn{4}{c}{density met, $N/N_0$} & energy per molecule, \\
\cmidrule(lr){2-5}
model & pulse 2 & pulse 3 & pulse 4 & pulse 5 & pulse 5 (eV) \\
\midrule
\multicolumn{6}{l}{\textit{200~ns between pulses (5~MHz)}} \\
coupled model & 0.870 & 0.673 & 0.517 & 0.393 & 0.340 \\
constant-volume & 1.007 & 1.014 & 1.021 & 1.027 & 0.128 \\
instantaneous-expansion & 0.812 & 0.649 & 0.512 & 0.399 & 0.334 \\
acoustic-relaxation & 0.907 & 0.779 & 0.648 & 0.528 & 0.252 \\
\addlinespace[4pt]
\multicolumn{6}{l}{\textit{430~ns between pulses (2.3~MHz)}} \\
coupled model & 0.769 & 0.599 & 0.463 & 0.353 & 0.378 \\
constant-volume & 1.007 & 1.014 & 1.020 & 1.026 & 0.128 \\
instantaneous-expansion & 0.808 & 0.643 & 0.505 & 0.389 & 0.342 \\
acoustic-relaxation & 0.850 & 0.689 & 0.547 & 0.427 & 0.312 \\
\bottomrule
\end{tabular}}
\end{table}

The existing closures miss this evolution of the density in different ways, as table~\ref{tab:burst} shows for each pulse. The constant-volume closure keeps the density, and so the energy per molecule, nearly constant. The fifth pulse meets a gas 2.6--2.9 times denser than in the coupled model and receives 2.7--3.0 times less energy per molecule. The acoustic-relaxation closure expands too slowly, as in case~1: at the fifth pulse its density is 21--34\% above that of the coupled model. The instantaneous-expansion closure comes closest, but its error changes sign with the pulse interval: with 200~ns between pulses the second and third pulses arrive while the axis is still expanding, and its density is up to 7\% below that of the coupled model. With 430~ns each pulse arrives near the density minimum left by the previous one, below the plateau that the closure assumes, and its density is 5--10\% above that of the coupled model.

For repeated pulses of high repetition frequency, solving the gas dynamics with the plasma chemistry therefore changes the energy per molecule of the later pulses by up to a factor of three compared with the constant-volume closure, which is the closure most global models use. In addition, no closure follows the coupled model at both pulse intervals studied here.

\FloatBarrier
\section{Discussion and limitations}\label{sec:discussion}

\subsection{Regimes of applicability of the reduced model}\label{sec:disc-where}

The reduced model fills a gap between the closures of global models and CFD simulations: within its scope it stays within a few percent of the reference CFD solution, where the closures miss by tens of percent, for about a thousandth of the computing time and with no fitted physical coefficient. Coupled to detailed kinetics, it opens lines of research that global models could not follow before. In plasma-assisted combustion, ignition by repetitive pulses depends on how successive pulses interact through the heat, the radicals and the flow left by the previous ones \cite{LefkowitzOmbrello2017,BechaneFiorina2024}. At megahertz frequencies the expansion adds to this interaction through the density that each pulse meets, and the model makes this part computable with detailed chemistry at a cost far below that of a CFD simulation. The gas dynamics takes less than a second, and the coupled calculation integrates the kinetics in a few tens of runs (41 for the single pulse of case~3; 31 for the bursts of case~4, the 30 shells that receive energy and the axis), once per iteration for two to five iterations, where a CFD calculation on the mesh of section~\ref{sec:reference} would integrate it in each of its 1000--2000 cells. Low-Mach large-eddy simulations of burst ignition cannot represent the acoustic expansion. They remove a fixed part of the deposited energy, set from separate compressible computations \cite{BechaneFiorina2024}, or impose the state at the end of the expansion \cite{roger2026modeling}, which does not hold when the next pulse arrives during the expansion. The proposed reduced model could supply the density during this phase.

Based on the obtained results, the model should replace a closure whenever a global model follows the gas during the expansion phase: in the first microseconds after every pulse, and for any deposition of energy during this phase (section~\ref{sec:intro}). The window within which no closure stays within 5\% of the reference axis density (for $\pk \ge 1.3$), 0.2 to $2.5\,\tac$, scales with the kernel radius and the sound speed. In air it spans about 0.06--0.7~\textmu s after the pulse for a thin channel ($r_0 = 100$~\textmu m at 300~K) and for the preheated spark of Rusterholtz et al.\ (\ref{app:rusterholtz}; $r_0 = 225$~\textmu m at 1500~K), 0.26--3.2~\textmu s for the kernel of case~3 ($r_0 = 450$~\textmu m) and 0.6--7~\textmu s for a kernel of 1~mm. With 10~\textmu s or more between pulses (100~kHz or less), the next pulse meets the gas after the expansion phase of kernels up to about a millimetre. The density it meets is then set by the isentropic plateau, the slow heat release, heat conduction and the flow \cite{Wang2022}, and the pulses are coupled by the heat and the radicals left by the previous pulse \cite{LefkowitzOmbrello2017}. The reduced model is then needed only for the first microseconds after each pulse.

\subsection{Comparison with published measurements}\label{sec:disc-measurements}

\ref{app:measurements} compares the models with two published experiments in air. Neither measured the deposited energy or its profile, which are necessary inputs of every model, so the comparisons can only test the coupled model as a whole, and not the reduced gas-dynamic model alone. Nevertheless, they provide a preliminary validation. In summary, the gas density measured by Lo et al.\ during the expansion phase is closer to the coupled model (19\% on average) than to any closure (31\% or more), although the coupled model is still up to 36\% too high. The atomic oxygen measured by Rusterholtz et al.\ before the gas moves agrees with our kinetics and rules out the instantaneous-expansion closure. A direct test of the gas dynamics needs the deposited energy, its profile and the axis density measured together. Density measurements without them, such as those of Wang et al.\ \cite{Wang2021a}, cannot provide this test, so no further validation is possible at this stage.

\subsection{Limitations}\label{sec:disc-limits}

The results reported here are limited to a straight channel at mid-gap, where the flow is cylindrical and one-dimensional. This holds for gaps longer than a few millimetres, where the channel does not collapse into a torus \cite{Dumitrache2019}, and until axial flow from the electrodes or the vortex rings they induce reach the measurement point, after several microseconds to tens of microseconds \cite{Singh2020}. The first microseconds after a pulse, and pulses repeated within them, end well before that. The measurement point of Lo et al.\ lies 1.5~mm from the plane electrode, and an order-of-magnitude estimate of the axial pressure gradient there makes axial flow unlikely to matter within the first microsecond, so the one-dimensional description is expected to hold over the compared times.

In addition, the reduced model is verified for smooth deposition profiles, from Gaussian to super-Gaussian of order about 4 and exponential, and up to $\pk \approx 6$ (within about 3\% at $\pk = 10$). Sharp (top-hat) profiles and strong blasts, whose shocks are not weak inside $R_\mathrm{out}$, are outside its scope, and the model is not meant for them. The deposition profile is an input, and so is the energy that heats the gas. In a sensitivity study with the reduced model at $\pk = 2$, in which we varied one input at a time, changing the profile from Gaussian to super-Gaussian of order 4 changed the axis density by up to 15\% within one acoustic time, more than the error of the model and less than the error of the closures. At later times the slow heat release matters more. Releasing 25--50\% of the deposited energy with time constants of 1--$5\,\tac$ changed the axis density by up to 18\%. In the coupled model the slow heat release is supplied by the kinetics, and it is itself uncertain. At the conditions of Lo et al., our kinetics releases on the axis about five times less heat between 0.15 and 1~\textmu s than their measured temperature implies.

Finally, we note two assumptions of the coupled calculations. First, the gas dynamics uses one value of $\gamma$, while the kinetics carries the real thermodynamics, dissociation and vibrational relaxation. Because the shells receive the entropy history computed by the kinetics along their own volume history, the converged solution depends on that value by less than 0.02\% (1.40 against 1.45, one case). Second, the electron-impact tables are computed at 300~K for the initial composition, without the vibrational excitation and the atomic oxygen that the pulses produce (7.6\% O on the axis at the fifth pulse with 200~ns between pulses), and the axis reaches about 1050--1170~K by the end of the bursts of case~4. We did not quantify the resulting error. A direct remedy is to solve the Boltzmann equation during the run, as in \ref{app:rusterholtz}; our global solver does this in the local-field approximation, but not yet with the LMEA used for the coupled calculations, and a detailed analysis of this effect is left for a future study.

\FloatBarrier
\section{Conclusions}\label{sec:conclusions}

We considered the gas density on the axis of a nanosecond discharge kernel in air during the expansion phase that follows the pulse: the density that sets the chemistry of the first microseconds and the reduced field of any later pulse. Compared with verified reference simulations of the compressible Euler equations, the closures that global kinetic models use for the expansion miss the axis density by tens of percent within the first few acoustic times, and the constant-volume closure by up to 151\% for a kernel heated to three times the ambient pressure. The existing closures miss two effects: a) the expansion takes a time set by the width of the deposition and the sound speed, and b) the inertia of the expanding gas carries the axis density below its final value before it recovers.

The reduced model proposed here, based on Lagrangian shells closed by an approximation of the cylindrical acoustic radiation condition, represents both effects, with no fitted physical coefficient. For smooth deposition profiles and pressure ratios up to about $\pk = 6$ it reproduces the reference axis density within a few percent, for single and repeated pulses and with a slow heat release, for about a thousandth of the computing time of the reference solution.

Coupled to a global plasma-kinetics solver (the coupled model), the reduced model shows that the choice of closure matters after every pulse. Two microseconds after a single pulse, and compared with the coupled model, the closures overestimate the axis density and the atomic oxygen by up to a factor of about three and the temperature by several hundred kelvin. When pulses follow each other during the expansion phase, each pulse meets a lower density and deposits more energy per molecule. No existing closure follows this growth at both pulse intervals studied. The proposed model should be used in a global model in two situations: in the first microseconds after every pulse, at any repetition rate, and whenever energy is deposited again during the expansion phase, by secondary discharges, reflections, double pulses or megahertz bursts.

The reduced model thus fills an important gap between the existing closures of global models and CFD simulations. It makes it possible to include the gas expansion in global models of nanosecond pulsed discharges with detailed chemistry, for plasma-assisted combustion and other applications. The reduced model is not specific to air: its gas dynamics needs only the ratio of specific heats and the sound speed of the ambient gas (verified here for $\gamma = 1.40$ and 1.31), while the composition and the thermodynamics of the mixture enter through the entropy history computed by the kinetics (section~\ref{sec:coupling}). It can therefore be applied to other mixtures, such as NH$_3$/air, H$_2$/air or CO$_2$, with adequate reaction mechanisms and electron cross sections. Future work should focus on the validation of the model against an experiment that measures the deposition profile and the axis density together. In addition, we aim to extend the model to stronger kernels with a shock-capturing term in the shells, so that a shock adds its entropy to the gas it crosses, and to integrate the model into the global solver, so that the shells and their kinetics are solved together without outer iterations.

\section*{Acknowledgements}
This work was partially supported by the European Union Project CAIPIRINH3A, under the GA Number
101191768. Views and opinions expressed are however those of the author(s) only and do not
necessarily reflect those of the European Union or CINEA. Neither the European Union nor the
granting authority can be held responsible for them. The work of K. Kourtzanidis is also partially supported by the H.F.R.I. project CO2-SPLIT. The
research project is implemented in the framework of H.F.R.I.
call ``4th Call for H.F.R.I.'s Research Projects to Support Postdoctoral Researchers'' (H.F.R.I. Project Number: 28223). For the purpose of open access, the authors have applied a Creative Commons Attribution (CC BY) licence to any Author Accepted Manuscript version arising from this submission.

\section*{Data availability}

The reaction mechanism used in this work (species, reactions, rate coefficients, energy partition and their sources) and the comparison of SoEEDF with BOLSIG+ are provided as supplementary material. The other data that support the findings of this study are available from the corresponding author upon reasonable request. The global solver of SoPlasma and the electron Boltzmann solver SoEEDF are proprietary and are not released. The reduced model is fully specified in section~\ref{sec:reduced} and \ref{app:boundary}.

\section*{Declaration of generative AI and AI-assisted technologies in the manuscript preparation process}
During the preparation of this work the authors used Claude (Anthropic) to assist with literature
summaries, language editing, data analysis and plotting scripts. After using this tool, the authors
reviewed and edited the content as needed and take full responsibility for the content of the
publication.

\appendix
\FloatBarrier
\section{The outer boundary condition of the reduced model}\label{app:boundary}

The gas outside $R_\mathrm{out}$ is replaced by a boundary condition on the outer face. In linear acoustics the outgoing cylindrical wave obeys a radiation condition with memory \cite{Diebold1991}: for each Fourier mode it is a convolution in time whose Laplace transform is a ratio of modified Bessel functions, $K_0/K_1$ for the axisymmetric mode, with $K_0$ and $K_1$ the modified Bessel functions of the second kind of orders zero and one, which describe a cylindrical wave that decays away from its source. For a spherical wave, in three dimensions, the corresponding ratio is a ratio of polynomials in the Laplace variable, so the radiation condition becomes a single ordinary differential equation, exact and local in time. For our cylindrical wave it is not, because $K_0/K_1$ contains a logarithm. For a small argument $x$, $K_0(x) \approx -\ln(x/2) - \gamma_\mathrm{E}$ and $K_1(x) \approx 1/x$, with $\gamma_\mathrm{E} \approx 0.577$ the Euler constant, so that $K_0(x)/K_1(x) \approx -x\,[\ln(x/2) + \gamma_\mathrm{E}]$, which no ratio of polynomials can follow, and the exact condition is not reproduced by any finite set of ordinary differential equations \cite{Alpert1998}. It can, however, be approximated by a sum of a few simple fractions (poles), and each fraction turns into one auxiliary ordinary differential equation \cite{Alpert1998}.

We approximate it by a sum of four poles, equation~(\ref{eq:poles}). We obtained the pole positions and weights $(t_k, a_k)$ of section~\ref{sec:reduced} by minimising the largest relative error of $\sum_k a_k x/(x + t_k)$ against $K_0(x)/K_1(x)$ on the imaginary axis for $0.003 \le |x| \le 300$, with real negative poles and non-negative weights that add up to one, so that the plane-wave limit is exact and the linearised boundary is passive. The largest error there is 1.4\%. Lower frequencies only matter after about $300\,R_\mathrm{out}/c_\mathrm{a}$, beyond every time window of this work. Alpert et al.\ approximate the reciprocal ratio with many complex poles to $10^{-6}$ accuracy. Our few-pole, non-negative fit is coarser and is judged here by the error of the whole model.

Four poles are enough for this work. For a weak kernel ($\pk = 1.1$) the outgoing wave is nearly linear, so the boundary condition reduces to the four-pole filter, and the whole model reproduces the reference axis density within 0.05\% (table~\ref{tab:verification}). At larger pressure ratios the remaining error comes from the nonlinear outgoing wave at the boundary, not from the fit of the poles: at $\pk = 6$ with $R_\mathrm{out} = 3 r_0$ it stays at about 8\% whether 20 or 40 shells are used, and it falls to 0.54\% when the boundary is moved to $8 r_0$, where the wave is weaker.

The filtered velocity $w$ is mapped to the boundary pressure by the plane simple wave of a polytropic gas \cite{Whitham1999}, obtained by holding the Riemann invariant of the incoming characteristic at its ambient value and eliminating the density with the isentrope. This relation is exact for a plane simple wave of any amplitude as long as no shock forms, while the pole filter is exact, to the accuracy of the fit, for a cylindrical wave of small amplitude. We combine the two by applying the plane-wave relation to the filtered velocity $w$, which carries the cylindrical correction, and obtain equation~(\ref{eq:simplewave}). Neither limit holds exactly for a strong cylindrical wave, so this combination is the main approximation of the boundary condition, besides the fit of the poles, and we judge it by the error of the whole model against the reference solution. The simpler, linear form of the same boundary condition, $p_\mathrm{out} - p_\mathrm{a} = \rho_\mathrm{a} c_\mathrm{a} w$, which is the small-amplitude limit of equation~(\ref{eq:simplewave}) and has nothing to do with the closures of section~\ref{sec:closures}, leaves an error that does not decrease with $N$ (2.5\% at $\pk = 2$), because the outgoing wave is not weak at $R_\mathrm{out}$ (its velocity reaches about a quarter of the sound speed there at $\pk = 3$). Equation~(\ref{eq:simplewave}) reduces it to 0.1--0.5\%. The nonlinear boundary relation is therefore what keeps a small domain, of three kernel radii, accurate at $\pk$ = 2--3; stronger kernels still need a larger domain (section~\ref{sec:res-verification}).

Mass is conserved exactly. The semi-discrete system conserves the total energy, including the work done on the outer boundary, and the integrated budget closes to $10^{-9}$ of the deposited energy.

The rule for the settings of section~\ref{sec:reduced}, to enlarge the domain and add shells until the axis density stops changing, can be checked on our cases. In the four cases computed here in which $R_\mathrm{out} = 3 r_0$ was too small (errors of 4.6 to 22\%), the next setting we used, $R_\mathrm{out} = 8 r_0$ with the same or twice the number of shells, differs from it by at least 3.7\%, so this check alone would have rejected the small domain. This holds for an enlargement by a factor of about three; we have not tested smaller steps. The rule does not detect the error of the model itself, which grows above $\pk \approx 6$ (section~\ref{sec:res-verification}).

\FloatBarrier
\section{Comparison with published measurements}\label{app:measurements}
\setcounter{figure}{0}

We compare the models with two published experiments in air. Neither measured the energy deposited on the axis or its radial profile, which are inputs of every model. The comparisons therefore test the coupled model as a whole (inputs, kinetics and gas dynamics), and not the gas dynamics alone, which case~2 has already verified. We present these comparisons as a preliminary validation. A full validation of the reduced model needs the deposited energy, its profile and the axis density measured together.

\subsection{Gas density during the expansion phase}\label{app:lo}

\begin{figure}[!htbp]
\centering
\includegraphics[width=\textwidth]{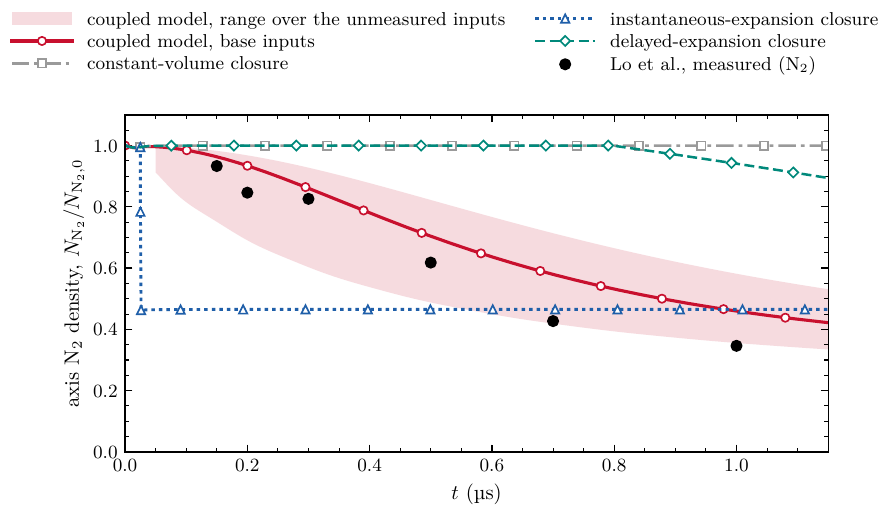}
\caption{Axis N$_2$ density after one pulse at the conditions of case~3, relative to its value before the pulse: measurements of Lo et al.\ \cite{Lo2014b}, the coupled model (line) and its range when the unmeasured inputs are varied (band), and the three closures of case~3.}
\label{fig:lo}
\end{figure}

Lo et al.\ \cite{Lo2014a,Lo2014b} measured the N$_2$ density on the axis of the discharge of case~3 from 0.15 to 1~\textmu s after the pulse. Less than 0.2\% of the N$_2$ is dissociated (supplementary material), so its density follows the gas density. We use the measurement at 0.15~\textmu s only to set the width of the deposition and the axis energy (case~3). The band in figure~\ref{fig:lo} shows how the prediction changes with the inputs that are not known exactly: the shape and width of the profile ($r_0 = 650$~\textmu m or an exponential profile) and the time origin of the measurements (shifted by 25~ns).

The coupled model is the closest to the measurements: its density differs from the measured one by 19\% on average, against 31\% for the instantaneous-expansion closure and 80--85\% for the other two. It is, however, too high at every time, by 4 to 36\%. This is not due to the heat release of our kinetics: with a heat release that makes the axis temperature follow the measured one, the computed density is still 7 to 35\% too high. We could not find the cause of this discrepancy.

\subsection{Temperature and atomic oxygen before the gas moves}\label{app:rusterholtz}

Rusterholtz et al.\ \cite{Rusterholtz2013} measured the gas temperature and the O($^3$P) density during and just after one pulse of a spark discharge in air preheated to 1500~K. At the compared times, 20 and 22.5~ns, the gas has not yet moved: these times correspond to less than 0.08 acoustic times, and the reference CFD solution of case~1 changes the axis density by only about 1\% by then. The coupled model then equals the constant-volume closure, so this measurement can only test our kinetics (reaction mechanism, cross sections and global solver) and the instantaneous-expansion closure. We use the inputs of Shao et al.\ \cite{Shao2024} for their code ChemPlasKin: the measured $E/N(t)$, and $n_\mathrm{e}(t)$ from a one-dimensional simulation, as in Cheng et al.\ \cite{Cheng2022}. With these inputs, our solver deposits 1.30~eV per molecule.

\begin{figure}[!htbp]
\centering
\includegraphics[width=\textwidth]{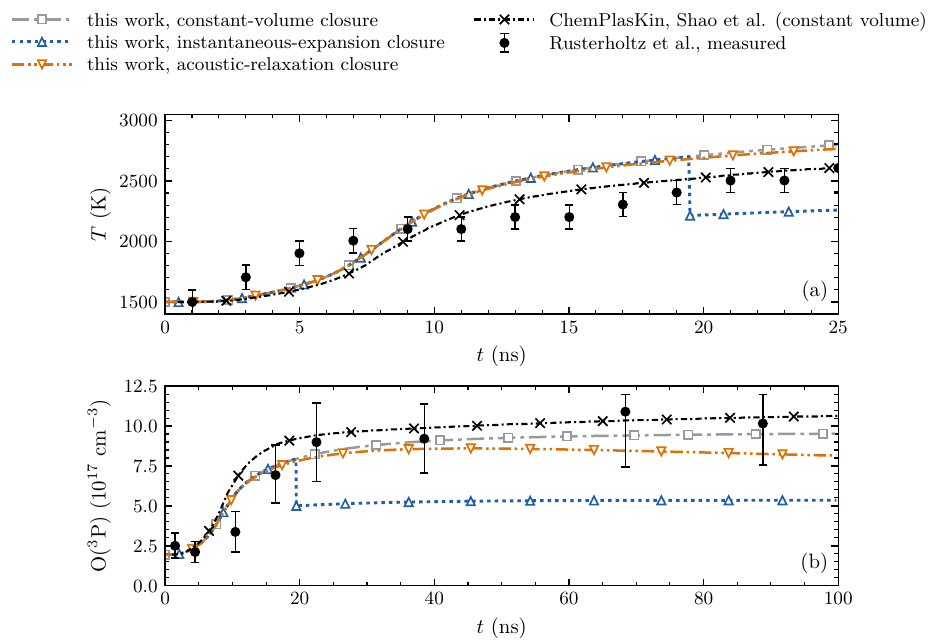}
\caption{(a) Gas temperature and (b) O($^3$P) density during and after one pulse of the spark of Rusterholtz et al.\ \cite{Rusterholtz2013}: measurements, our global solver with three closures, and ChemPlasKin (Shao et al.\ \cite{Shao2024}, at constant volume), replotted from Shao et al.}
\label{fig:rusterholtz}
\end{figure}

Our calculations show an O($^3$P) density at 22.5~ns which is 8\% below the measurement, while the calculations of ChemPlasKin show 5\% above. Both are in good agreement with the measurements, within the error bar of $\pm 27$\%. The instantaneous-expansion closure, which dilutes the gas at the end of the pulse, is 44\% below. Our calculations show a temperature at 20~ns of 2710~K, 11\% above the measured $2450 \pm 100$~K, while ChemPlasKin gives 2530~K which lies inside the experimental error bar. To understand this discrepancy, we compared our heating fraction with that of Popov for this discharge: 23\% of the deposited energy within 35~ns, against 24\% \cite{Popov2016}. The excess heat therefore comes from the deposited energy. With our heating fraction, the measured temperature needs about 1.03~eV per molecule, inside the 0.6--1.4~eV that the measured pulse energy allows, while the published inputs deposit 1.30~eV. Overall, the comparisons with the measurements and with independent calculations support our kinetics, and they rule out the instantaneous-expansion closure in this discharge.

\clearpage
\newgeometry{margin=1.8cm}
\begingroup\footnotesize\setlength{\parindent}{0pt}\setlength{\parskip}{4pt}\setlength{\tabcolsep}{4pt}
\section*{Supplementary material}
\addcontentsline{toc}{section}{Supplementary material}
\section*{The air reaction mechanism}

The global (zero-dimensional) calculations of the paper use the air mechanism listed here: 17 species
(Table~S1), 24 electron-impact reactions (Table~S2) and 38 heavy-particle reactions (Table~S3), with 27 further
electron-impact processes that enter only the electron Boltzmann equation (Table~S4). The cross sections are taken from
the LXCat platform \cite{Pitchford2017}: the Phelps database \cite{LXCatPhelps} for N$_2$ and O$_2$ and the IST-Lisbon
database \cite{LXCatISTLisbon} for ground-state N and O. They are not reproduced here; they are available from LXCat
(www.lxcat.net) under the databases cited. Electron-impact rate coefficients are computed from them with a two-term
electron Boltzmann solver, as described in the paper.

Heavy-particle rate coefficients are $k = A\,T^b e^{-E_a/T}$ with $T$ in K: the electron temperature for reactions
with an electron reactant, and the gas temperature otherwise. $\Delta H$ is the energy released by the process: the heat
of reaction for a heavy-particle reaction, and for electron impact the part of the electron energy that the products do
not store. A part $Q_\mathrm{p}$ of it heats the gas at once (fast gas heating). The rest, $\Delta H - Q_\mathrm{p}$,
is stored as given in the column ``rest stored as'': ``vibration'' is vibrational energy of N$_2$ and O$_2$, released
later as heat by vibrational--translational relaxation; ``not released'' is energy that does not heat the gas within the
times computed (see Assumptions and known gaps). Where no source gives a partition, the whole of $\Delta H$ heats the gas
at once. The thermodynamic data are NASA polynomials (Cantera, nasa\_gas.yaml); an excited state has the data of its
ground state with the enthalpy raised by its excitation energy. Assumptions and known gaps are listed at the end.

\section*{Table S1. Species}
\begin{longtable}{@{}l l r r@{}}
\toprule No. & species & $M$ (g\,mol$^{-1}$) & excitation energy (eV) \\ \midrule\endhead
\bottomrule\endfoot
1.1 & e & 0.00054858 & -- \\
1.2 & N$_2$ & 28.014 & -- \\
1.3 & O$_2$ & 31.998 & -- \\
1.4 & O & 15.999 & -- \\
1.5 & N & 14.007 & -- \\
1.6 & N$_2$(A$^3\Sigma_u^+$) & 28.014 & 6.17 \\
1.7 & N$_2$(B$^3\Pi_g$) & 28.014 & 7.35 \\
1.8 & N$_2$(a$'^1\Sigma_u^-$) & 28.014 & 8.4 \\
1.9 & O($^1$D) & 15.999 & 1.96 \\
1.10 & O($^1$S) & 15.999 & 4.19 \\
1.11 & N$_2$(C$^3\Pi_u$) & 28.014 & 11.03 \\
1.12 & O$_2$(a$^1\Delta_g$) & 31.998 & 0.977 \\
1.13 & O$_2$(b$^1\Sigma_g^+$) & 31.998 & 1.627 \\
1.14 & N$_2^+$ & 28.013 & -- \\
1.15 & O$_2^+$ & 31.997 & -- \\
1.16 & O$^-$ & 16 & -- \\
1.17 & O$_2^-$ & 31.999 & -- \\
\end{longtable}

\section*{Table S2. Electron-impact reactions}
Rate coefficients from the cross sections cited, integrated over the electron energy distribution. The threshold is given after each reaction.
\begin{longtable}{@{}l p{0.34\linewidth} r r l p{0.25\linewidth}@{}}
\toprule No. & reaction (threshold) & $\Delta H$ (eV) & $Q_\mathrm{p}$ (eV) & rest stored as & cross section \\ \midrule\endhead
\bottomrule\endfoot
\multicolumn{6}{@{}l}{\textit{electron impact: electronic excitation}} \\
2.1 & e + N$_2$ $\rightarrow$ N$_2$(A$^3\Sigma_u^+$) + e (6.17\,eV) & 0 & 0 & -- & \cite{LXCatPhelps,Phelps1985} \\
2.2 & e + N$_2$ $\rightarrow$ N$_2$(A$^3\Sigma_u^+$) + e (7\,eV) & 0.83 & 0 & vibration & \cite{LXCatPhelps,Phelps1985} \\
2.3 & e + N$_2$ $\rightarrow$ N$_2$(B$^3\Pi_g$) + e (7.35\,eV) & 0 & 0 & -- & \cite{LXCatPhelps,Phelps1985} \\
2.4 & e + N$_2$ $\rightarrow$ N$_2$(B$^3\Pi_g$) + e (7.36\,eV) & 0 & 0 & -- & \cite{LXCatPhelps,Phelps1985} \\
2.5 & e + N$_2$ $\rightarrow$ N$_2$(A$^3\Sigma_u^+$) + e (7.8\,eV) & 1.63 & 0 & vibration & \cite{LXCatPhelps,Phelps1985} \\
2.6 & e + N$_2$ $\rightarrow$ N$_2$(B$^3\Pi_g$) + e (8.16\,eV) & 0.81 & 0 & vibration & \cite{LXCatPhelps,Phelps1985} \\
2.7 & e + N$_2$ $\rightarrow$ N$_2$(a$'^1\Sigma_u^-$) + e (8.4\,eV) & 0 & 0 & -- & \cite{LXCatPhelps,Phelps1985} \\
2.8 & e + N$_2$ $\rightarrow$ N$_2$(a$'^1\Sigma_u^-$) + e (8.55\,eV) & 0.15 & 0 & vibration & \cite{LXCatPhelps,Phelps1985} \\
2.9 & e + N$_2$ $\rightarrow$ N$_2$(a$'^1\Sigma_u^-$) + e (8.89\,eV) & 0.49 & 0 & vibration & \cite{LXCatPhelps,Phelps1985} \\
2.10 & e + N$_2$ $\rightarrow$ N$_2$(C$^3\Pi_u$) + e (11.03\,eV) & 0 & 0 & -- & \cite{LXCatPhelps,Phelps1985} \\
2.11 & e + N$_2$ $\rightarrow$ N$_2$(C$^3\Pi_u$) + e (11.87\,eV) & 0.84 & 0 & vibration & \cite{LXCatPhelps,Phelps1985} \\
2.12 & e + N$_2$ $\rightarrow$ N$_2$(C$^3\Pi_u$) + e (12.25\,eV) & 1.22 & 0 & vibration & \cite{LXCatPhelps,Phelps1985} \\
2.13 & e + O$_2$ $\rightarrow$ O$_2$(a$^1\Delta_g$) + e (0.977\,eV) & 0 & 0 & -- & \cite{LXCatPhelps,Lawton1978} \\
2.14 & e + O$_2$ $\rightarrow$ O$_2$(b$^1\Sigma_g^+$) + e (1.627\,eV) & 0 & 0 & -- & \cite{LXCatPhelps,Lawton1978} \\
2.15 & e + O $\rightarrow$ O($^1$D) + e (1.96\,eV) & 0 & 0 & -- & \cite{LXCatISTLisbon,Laher1990} \\
2.16 & e + O $\rightarrow$ O($^1$S) + e (4.19\,eV) & 0 & 0 & -- & \cite{LXCatISTLisbon,Laher1990} \\
\multicolumn{6}{@{}l}{\textit{electron impact: dissociation}} \\
2.17 & e + O$_2$ $\rightarrow$ O + O + e (6\,eV) & 0.835 & 0.835 & -- & \cite{LXCatPhelps,Lawton1978} \\
2.18 & e + O$_2$ $\rightarrow$ O + O($^1$D) + e (8.4\,eV) & 1.275 & 1.275 & -- & \cite{LXCatPhelps,Lawton1978}; partition \cite{Popov2011,Mintoussov2011,PopovStarikovskaia2022} \\
2.19 & e + O$_2$ $\rightarrow$ O + O + e (9.97\,eV) & 4.805 & 4.805 & -- & \cite{LXCatPhelps} \\
\multicolumn{6}{@{}l}{\textit{electron impact: ionisation}} \\
2.20 & e + N$_2$ $\rightarrow$ N$_2^+$ + e + e (15.6\,eV) & 0 & 0 & -- & \cite{LXCatPhelps,Phelps1985} \\
2.21 & e + N$_2$ $\rightarrow$ N$_2^+$ + e + e (18.8\,eV) & 3.155 & 3.155 & -- & \cite{LXCatPhelps,Phelps1985} \\
2.22 & e + O$_2$ $\rightarrow$ O$_2^+$ + e + e (12.06\,eV) & 0 & 0 & -- & \cite{LXCatPhelps,Lawton1978} \\
\multicolumn{6}{@{}l}{\textit{electron impact: attachment}} \\
2.23 & e + O$_2$ $\rightarrow$ O$_2^-$ & 0 & 0 & -- & \cite{LXCatPhelps,Lawton1978} \\
2.24 & e + O$_2$ $\rightarrow$ O$^-$ + O & 0 & 0 & -- & \cite{LXCatPhelps,Lawton1978} \\
\end{longtable}

\begin{landscape}
\section*{Table S3. Heavy-particle reactions}
$k = A\,T^{b} e^{-E_a/T}$ in cm$^3$\,s$^{-1}$ ($^{\dagger}$cm$^6$\,s$^{-1}$, three-body), $T$ in K.
\begin{longtable}{@{}l p{0.25\linewidth} l r r l p{0.28\linewidth}@{}}
\toprule No. & reaction & $k$ & $\Delta H$ (eV) & $Q_\mathrm{p}$ (eV) & rest stored as & source \\ \midrule\endhead
\bottomrule\endfoot
\multicolumn{7}{@{}l}{\textit{electron-ion recombination}} \\
3.1 & e + N$_2^+$ $\rightarrow$ N + N & $2.035\times10^{-6}\,T^{-0.39}$ & 5.847 & 2.097 & vibration & \cite{SheehanStMaurice2004,MehrBiondi1969}; partition \cite{Peterson1998} \\
3.2 & e + O$_2^+$ $\rightarrow$ O + O & $1.277\times10^{-6}\,T^{-0.61}$ & 6.98 & 6.98 & -- & \cite{SheehanStMaurice2004,Peverall2001} (20.4\% branch) \\
3.3 & e + O$_2^+$ $\rightarrow$ O + O($^1$D) & $2.755\times10^{-6}\,T^{-0.61}$ & 5.02 & 5.02 & -- & \cite{SheehanStMaurice2004,Peverall2001} (44\% branch) \\
3.4 & e + O$_2^+$ $\rightarrow$ O($^1$D) + O($^1$D) & $1.972\times10^{-6}\,T^{-0.61}$ & 3.06 & 3.06 & -- & \cite{SheehanStMaurice2004,Peverall2001} (31.5\% branch) \\
3.5 & e + O$_2^+$ $\rightarrow$ O($^1$D) + O($^1$S) & $2.567\times10^{-7}\,T^{-0.61}$ & 0.8301 & 0.8301 & -- & \cite{SheehanStMaurice2004,Peverall2001} (4.1\% branch) \\
\multicolumn{7}{@{}l}{\textit{ion-ion recombination}} \\
3.6 & O$^-$ + O$_2^+$ $\rightarrow$ O + O$_2$ & $3.464\times10^{-6}\,T^{-0.5}$ & 10.62 & 10.62 & -- & \cite{Kossyi1992} \\
3.7 & O$^-$ + N$_2^+$ $\rightarrow$ O + N$_2$ & $3.464\times10^{-6}\,T^{-0.5}$ & 14.12 & 14.12 & -- & \cite{Kossyi1992} \\
3.8 & O$_2^-$ + O$_2^+$ $\rightarrow$ O$_2$ + O$_2$ & $3.464\times10^{-6}\,T^{-0.5}$ & 11.65 & 11.65 & -- & \cite{Kossyi1992} \\
3.9 & O$_2^-$ + N$_2^+$ $\rightarrow$ O$_2$ + N$_2$ & $3.464\times10^{-6}\,T^{-0.5}$ & 15.15 & 15.15 & -- & \cite{Kossyi1992} \\
\multicolumn{7}{@{}l}{\textit{detachment}} \\
3.10 & O$^-$ + O $\rightarrow$ O$_2$ + e & $5\times10^{-10}$ & 3.636 & 3.636 & -- & \cite{Kossyi1992} \\
3.11 & O$_2^-$ + O$_2$ $\rightarrow$ O$_2$ + O$_2$ + e & $1.559\times10^{-11}\,T^{0.5}\,e^{-5590/T}$ & -0.4978 & -0.4978 & -- & \cite{Kossyi1992} \\
\multicolumn{7}{@{}l}{\textit{ion-molecule reactions}} \\
3.12 & O$_2^-$ + O $\rightarrow$ O$_2$ + O$^-$ & $3.3\times10^{-10}$ & 1.031 & 1.031 & -- & \cite{Kossyi1992} \\
\multicolumn{7}{@{}l}{\textit{quenching and energy transfer of excited states}} \\
3.13 & N$_2$(A$^3\Sigma_u^+$) + O$_2$ $\rightarrow$ N$_2$ + O + O & $2.54\times10^{-12}$ & 1.005 & 1.005 & -- & \cite{Kossyi1992} \\
3.14 & N$_2$(A$^3\Sigma_u^+$) + O $\rightarrow$ N$_2$ + O($^1$S) & $2.1\times10^{-11}$ & 1.98 & 0 & vibration & \cite{Kossyi1992,Popov2016}; partition \cite{Popov2016} \\
3.15 & N$_2$(A$^3\Sigma_u^+$) + N$_2$ $\rightarrow$ N$_2$ + N$_2$ & $3\times10^{-18}$ & 6.17 & 6.17 & -- & \cite{Kossyi1992} \\
3.16 & N$_2$(B$^3\Pi_g$) + N$_2$ $\rightarrow$ N$_2$(A$^3\Sigma_u^+$) + N$_2$ & $5\times10^{-11}$ & 1.18 & 0 & vibration & \cite{Kossyi1992}; partition \cite{Popov2016} \\
3.17 & N$_2$(B$^3\Pi_g$) + O$_2$ $\rightarrow$ N$_2$ + O + O & $3\times10^{-10}$ & 2.185 & 2.185 & -- & \cite{Kossyi1992,Rusterholtz2013} \\
3.18 & N$_2$(a$'^1\Sigma_u^-$) + O$_2$ $\rightarrow$ N$_2$ + O + O($^1$D) & $4.3\times10^{-10}$ & 1.275 & 1.275 & -- & \cite{Marinelli1989,Popov2011} \\
3.19 & O($^1$D) + N$_2$ $\rightarrow$ O + N$_2$ & $1.8\times10^{-11}\,e^{107/T}$ & 1.96 & 1.38 & vibration & \cite{Mintoussov2011,Popov2016}; partition \cite{Popov2016,Mintoussov2011} \\
3.20 & O($^1$D) + O$_2$ $\rightarrow$ O + O$_2$(b$^1\Sigma_g^+$) & $2.56\times10^{-11}\,e^{67/T}$ & 0.333 & 0 & vibration & \cite{Mintoussov2011} (80\% branch); partition \cite{Popov2016} \\
3.21 & O($^1$D) + O$_2$ $\rightarrow$ O + O$_2$ & $6.4\times10^{-12}\,e^{67/T}$ & 1.96 & 0 & vibration & \cite{Mintoussov2011} (20\% branch); partition \cite{Popov2016} \\
3.22 & N$_2$(a$'^1\Sigma_u^-$) + N$_2$ $\rightarrow$ N$_2$(B$^3\Pi_g$) + N$_2$ & $2\times10^{-13}$ & 1.05 & 0 & vibration & \cite{Kossyi1992}; partition \cite{Popov2016} \\
3.23 & N$_2$(C$^3\Pi_u$) + N$_2$ $\rightarrow$ N$_2$(a$'^1\Sigma_u^-$) + N$_2$ & $10^{-11}$ & 2.63 & 0 & vibration & \cite{Kossyi1992}; partition \cite{Popov2016} \\
3.24 & N$_2$(C$^3\Pi_u$) + O$_2$ $\rightarrow$ N$_2$ + O + O & $2.492\times10^{-11}\,T^{0.3}$ & 5.865 & 5.554 & vibration & \cite{Rusterholtz2013,Cheng2022} (46\% branch); partition \cite{Cheng2022} \\
3.25 & N$_2$(C$^3\Pi_u$) + O$_2$ $\rightarrow$ N$_2$ + O + O($^1$D) & $2.818\times10^{-11}\,T^{0.3}$ & 3.905 & 3.587 & vibration & \cite{Rusterholtz2013,Cheng2022} (52\% branch); partition \cite{Cheng2022} \\
3.26 & O$_2$(a$^1\Delta_g$) + O$_2$ $\rightarrow$ O$_2$ + O$_2$ & $2.295\times10^{-20}\,T^{0.8}$ & 0.977 & 0.977 & -- & \cite{Kossyi1992} \\
3.27 & O$_2$(a$^1\Delta_g$) + N$_2$ $\rightarrow$ O$_2$ + N$_2$ & $3\times10^{-21}$ & 0.977 & 0.977 & -- & \cite{Kossyi1992} \\
3.28 & O$_2$(b$^1\Sigma_g^+$) + O$_2$ $\rightarrow$ O$_2$(a$^1\Delta_g$) + O$_2$ & $4.3\times10^{-22}\,T^{2.4}\,e^{-241/T}$ & 0.65 & 0.65 & -- & \cite{Kossyi1992} \\
3.29 & O$_2$(b$^1\Sigma_g^+$) + N$_2$ $\rightarrow$ O$_2$(a$^1\Delta_g$) + N$_2$ & $4.9\times10^{-15}\,e^{-253/T}$ & 0.65 & 0.65 & -- & \cite{Kossyi1992} \\
3.30 & N$_2$(A$^3\Sigma_u^+$) + N$_2$(A$^3\Sigma_u^+$) $\rightarrow$ N$_2$ + N$_2$(B$^3\Pi_g$) & $7.7\times10^{-11}$ & 4.99 & 3.5 & vibration & \cite{PopovStarikovskaia2022,Popov2011}; partition \cite{Popov2011} \\
3.31 & O($^1$S) + O $\rightarrow$ O($^1$D) + O & $5\times10^{-11}\,e^{-301/T}$ & 2.23 & 2.23 & -- & \cite{Popov2016}; partition \cite{Popov2016} \\
3.32 & O($^1$S) + O$_2$ $\rightarrow$ O + O$_2$ & $3\times10^{-12}\,e^{-850/T}$ & 4.19 & 0 & vibration & \cite{Popov2016}; partition \cite{Popov2016} \\
3.33 & O($^1$S) + O$_2$ $\rightarrow$ O($^1$D) + O$_2$ & $1.3\times10^{-12}\,e^{-850/T}$ & 2.23 & 0 & vibration & \cite{Popov2016}; partition \cite{Popov2016} \\
3.34 & O($^1$S) + N$_2$ $\rightarrow$ O($^1$D) + N$_2$ & $5\times10^{-17}$ & 2.23 & 0 & vibration & \cite{Popov2016}; partition \cite{Popov2016} \\
3.35 & N$_2$(C$^3\Pi_u$) + O$_2$ $\rightarrow$ N$_2$ + O + O($^1$S) & $1.084\times10^{-12}\,T^{0.3}$ & 1.675 & 1.368 & vibration & \cite{Rusterholtz2013,Cheng2022} (2\% branch); partition \cite{Cheng2022} \\
3.36 & N$_2$(A$^3\Sigma_u^+$) + N$_2$(A$^3\Sigma_u^+$) $\rightarrow$ N$_2$ + N$_2$(C$^3\Pi_u$) & $1.6\times10^{-10}$ & 1.31 & 0 & vibration & \cite{PopovStarikovskaia2022}; partition \cite{Popov2011} \\
\multicolumn{7}{@{}l}{\textit{neutral chemistry}} \\
3.37 & O + O + N$_2$ $\rightarrow$ O$_2$ + N$_2$ & $2.76\times10^{-34}\,e^{720/T}$$^{\dagger}$ & 5.165 & 5.165 & -- & \cite{Kossyi1992} \\
3.38 & N + N + N$_2$ $\rightarrow$ N$_2$ + N$_2$ & $8.27\times10^{-34}\,e^{500/T}$$^{\dagger}$ & 9.798 & 9.798 & -- & \cite{Kossyi1992} \\
\end{longtable}
\end{landscape}

\section*{Table S4. Electron-impact processes in the electron Boltzmann equation only}
These processes shape the electron energy distribution and the energy partition, but change no species density. Elastic collisions (4.1--4.4) heat the gas through the elastic power, which the global solver computes separately.
\begin{longtable}{@{}l p{0.26\linewidth} r r l p{0.30\linewidth}@{}}
\toprule No. & process & threshold (eV) & $Q_\mathrm{p}$ (eV) & rest stored as & cross section \\ \midrule\endhead
\bottomrule\endfoot
\multicolumn{6}{@{}l}{\textit{electron impact: momentum transfer}} \\
4.1 & N$_2$ & -- & 0 & -- & \cite{LXCatPhelps,Phelps1985} \\
4.2 & O$_2$ & -- & 0 & -- & \cite{LXCatPhelps,Lawton1978} \\
4.3 & O & 0 & 0 & -- & \cite{LXCatISTLisbon} \\
4.4 & N & 0 & 0 & -- & \cite{LXCatISTLisbon,Wang2014} \\
\multicolumn{6}{@{}l}{\textit{electron impact: rotational excitation}} \\
4.5 & N$_2$ $\rightarrow$ N$_2$(rot) & 0.02 & 0 & -- & \cite{LXCatPhelps,Phelps1985} \\
4.6 & O$_2$ $\rightarrow$ O$_2$(rot) & 0.02 & 0 & -- & \cite{LXCatPhelps,Lawton1978} \\
\multicolumn{6}{@{}l}{\textit{electron impact: vibrational excitation}} \\
4.7 & N$_2$ $\rightarrow$ N$_2$($v$=1) & 0.29 & 0 & vibration & \cite{LXCatPhelps,Phelps1985} \\
4.8 & N$_2$ $\rightarrow$ N$_2$($v$=1, res) & 0.291 & 0 & vibration & \cite{LXCatPhelps,Phelps1985} \\
4.9 & N$_2$ $\rightarrow$ N$_2$($v$=2) & 0.59 & 0 & vibration & \cite{LXCatPhelps,Phelps1985} \\
4.10 & N$_2$ $\rightarrow$ N$_2$($v$=3) & 0.88 & 0 & vibration & \cite{LXCatPhelps,Phelps1985} \\
4.11 & N$_2$ $\rightarrow$ N$_2$($v$=4) & 1.17 & 0 & vibration & \cite{LXCatPhelps,Phelps1985} \\
4.12 & N$_2$ $\rightarrow$ N$_2$($v$=5) & 1.47 & 0 & vibration & \cite{LXCatPhelps,Phelps1985} \\
4.13 & N$_2$ $\rightarrow$ N$_2$($v$=6) & 1.76 & 0 & vibration & \cite{LXCatPhelps,Phelps1985} \\
4.14 & N$_2$ $\rightarrow$ N$_2$($v$=7) & 2.06 & 0 & vibration & \cite{LXCatPhelps,Phelps1985} \\
4.15 & N$_2$ $\rightarrow$ N$_2$($v$=8) & 2.35 & 0 & vibration & \cite{LXCatPhelps,Phelps1985} \\
4.16 & O$_2$ $\rightarrow$ O$_2$($v$=1) & 0.19 & 0 & vibration & \cite{LXCatPhelps,Lawton1978} \\
4.17 & O$_2$ $\rightarrow$ O$_2$($v$=1, res) & 0.19 & 0 & vibration & \cite{LXCatPhelps,Lawton1978} \\
4.18 & O$_2$ $\rightarrow$ O$_2$($v$=2) & 0.38 & 0 & vibration & \cite{LXCatPhelps,Lawton1978} \\
4.19 & O$_2$ $\rightarrow$ O$_2$($v$=2, res) & 0.38 & 0 & vibration & \cite{LXCatPhelps,Lawton1978} \\
4.20 & O$_2$ $\rightarrow$ O$_2$($v$=3) & 0.57 & 0 & vibration & \cite{LXCatPhelps,Lawton1978} \\
4.21 & O$_2$ $\rightarrow$ O$_2$($v$=4) & 0.75 & 0 & vibration & \cite{LXCatPhelps,Lawton1978} \\
\multicolumn{6}{@{}l}{\textit{electron impact: electronic excitation}} \\
4.22 & N$_2$ $\rightarrow$ N$_2$(SUM) & 13 & 3.25 & not released & \cite{LXCatPhelps,Phelps1985} \\
4.23 & O$_2$ $\rightarrow$ O$_2$(4.5 eV) & 4.5 & 4.5 & -- & \cite{LXCatPhelps,Lawton1978} \\
4.24 & N $\rightarrow$ N($^2$D) & 2.38 & 2.38 & -- & \cite{LXCatISTLisbon,Wang2014} \\
4.25 & N $\rightarrow$ N($^2$P) & 3.58 & 3.58 & -- & \cite{LXCatISTLisbon,Wang2014} \\
\multicolumn{6}{@{}l}{\textit{electron impact: ionisation}} \\
4.26 & O $\rightarrow$ O$^+$ & 13.62 & 0 & not released & \cite{LXCatISTLisbon,Laher1990} \\
4.27 & N $\rightarrow$ N$^+$ & 14.53 & 0 & not released & \cite{LXCatISTLisbon,Wang2014} \\
\end{longtable}

\section*{Assumptions and known gaps}
The energy partition of each process follows from its products or from the sources cited in Tables S2--S4, except in
the cases below. The shares of the electron power quoted are those computed by the electron Boltzmann solver for dry
air at the reduced fields of cases 3 and 4 of the paper (115--150~Td).

\textbf{Assumptions.}
\begin{itemize}\setlength{\itemsep}{1pt}
\item \emph{Lumped states} (2.6, 2.8, 2.9, 2.11, 2.12, 3.18). N$_2$ states that are not in the mechanism are counted as
the nearest state that is: B$'^3\Sigma_u^-$ as B$^3\Pi_g$; a$^1\Pi_g$ and w$^1\Delta_u$ as a$'^1\Sigma_u^-$;
E$^3\Sigma_g^+$ and a$''^1\Sigma_g^+$ as C$^3\Pi_u$. The energy difference goes to vibration. The quenching of
N$_2$(a$'^1\Sigma_u^-$) by O$_2$ (3.18) uses the rate measured for N$_2$(a$^1\Pi_g$).
\item \emph{Energy released as heat at once} (2.19, 2.21, 4.23--4.25). The energy of some states that are not in the
mechanism heats the gas at once: the O$_2$ Herzberg states at 4.5~eV (4.23), metastable states below the dissociation
limit of O$_2$ that are removed by collisions; N$_2^+$(B$^2\Sigma_u^+$) (2.21), which is quenched in well under a nanosecond at 1~atm, while its radiative
lifetime is about 62~ns; N($^2$D) and N($^2$P) (4.24, 4.25). In 2.19 the 4.8~eV above the O + O limit heats the gas,
although part of it is electronic energy of O($^1$D).
\item \emph{Energy not released} (4.22, 4.26, 4.27). The high-lying singlet states of N$_2$ (the sum of singlet states of
the Phelps database, at 13~eV) lie above the dissociation energy of N$_2$ (9.76~eV) and can dissociate it, but the
mechanism produces no N atoms from them: 9.75~eV is kept as dissociation energy and 3.25~eV heats the gas. The
ionisation of O and N atoms keeps its energy, since O$^+$ and N$^+$ are not in the mechanism.
\end{itemize}

\textbf{Known gaps.}
\begin{itemize}\setlength{\itemsep}{1pt}
\item \emph{Rotational excitation} (4.5, 4.6). Its energy is removed from the electrons but not added to the gas,
although rotational--translational relaxation turns it into heat within a few collisions.
\item \emph{Dissociative attachment} (2.24). The cross section has no threshold in the database, so no energy defect is
computed and the kinetic energy of the fragments does not heat the gas.
\end{itemize}

\textbf{Effect on the results.}
\begin{itemize}\setlength{\itemsep}{1pt}
\item Fast gas heating. Rotational excitation takes 0.6--1.0\% of the electron power and the energy difference of the
lumped states about 0.5\%; both may make the fast gas heating too low. The Herzberg states take 1.5\%, all released as
heat, which may make it too high. Each of these is 1.5\% of the deposited energy or less, small against the fast gas heating (15\%
of the deposited energy at the end of the pulse in case 3) and against the differences between our energy partition
and those of Lo et al.\ and Popov (up to 5\% of the deposited energy, Table 5 of the paper).
\item N$_2$ dissociation. The high-lying singlet states take 1.5--4\% of the electron power. If they all dissociated
N$_2$, about 0.15\% of the N$_2$ would be dissociated in case 3, in addition to the 0.02\% that the mechanism gives.
This does not change the gas density, and the reactions of these N atoms would release at most about 1.5\% of the
deposited energy as heat (the 9.75~eV kept per excitation).
\item Dissociative attachment (2.24), the 9.97~eV dissociation of O$_2$ (2.19) and N$_2^+$(B$^2\Sigma_u^+$) (2.21)
each take 0.2\% of the electron power or less, so their partition does not matter.
\item The processes on N and O atoms (4.24--4.27) do not enter the coupled calculations of the paper, whose electron
data are computed for N$_2$ and O$_2$ only; they enter only the comparison of Appendix B.2.
\end{itemize}

\section*{Verification of the electron Boltzmann solver}
Figure~S1 compares the mean electron energy, the reduced mobility and the reduced diffusion coefficient computed by
SoEEDF, the two-term electron Boltzmann solver of section~\ref{sec:soeedf}, with those of BOLSIG+ (version 07/2024) for dry air
(79\% N$_2$, 21\% O$_2$, Phelps cross sections, gas temperature 300~K, no electron--electron or superelastic
collisions, temporal growth), at 400 reduced fields from 1 to 1000~Td. BOLSIG+ is run on a converged grid of 1000
energy points. The largest deviation is 0.021\% (mean energy, at 64~Td); the mobility and the diffusion coefficient
agree within 0.011\%.

\begin{figure}[h]
\centering
\includegraphics{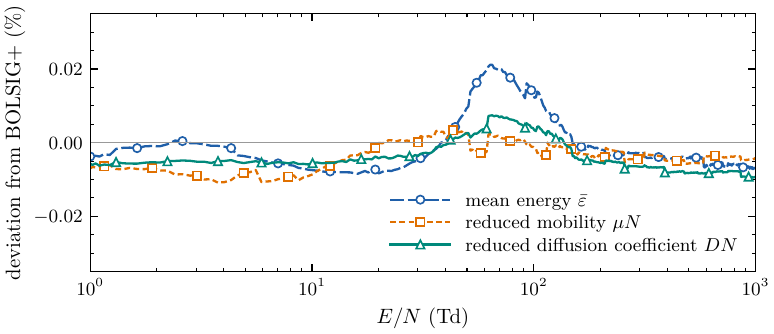}

{\small Figure S1. Relative deviation of SoEEDF from BOLSIG+ (converged 1000-point energy grid) for dry air at 300~K.}
\end{figure}
\endgroup

\end{document}